\documentclass[12pt]{article}
\usepackage{amsfonts}
\usepackage{amssymb}
\usepackage{amsbsy}
\usepackage{graphics}
\usepackage{epsfig}
\usepackage{color}
\usepackage{mathrsfs}
\usepackage{amsmath}

\input epsf.sty

\newlength{\TZ}
\newcommand{\BEQ}{\begin{equation}}     
\newcommand{\BEA}{\begin{eqnarray}}
\newcommand{\BD}{\begin{displaymath}}
\newcommand{\EEQ}{\end{equation}}       
\newcommand{\EEA}{\end{eqnarray}}
\newcommand{\ED}{\end{displaymath}}
\newcommand{\bb}{\begin{eqnarray}}
\newcommand{\ee}{\end{eqnarray}}

\newcommand{\e}{{\rm e}}

\newcommand{\D}{{\rm d}}                
\newcommand{\II}{{\rm i}}               
\renewcommand{\Re}{{\rm Re\ }}          
\renewcommand{\Im}{{\rm Im\ }}          
 
\newcommand{\demi}{\frac{1}{2}}         
\newcommand{\wht}[1]{\widehat{#1}}      
\newcommand{\bra}[1]{\left\langle#1\right|}  
\newcommand{\ket}[1]{\left|#1\right\rangle}  
\newcommand{\hp}{\wht{p}}
\newcommand{\hs}{\wht{s}}
\newcommand{\dicht}{\wht{\rho}}
\newcommand{\up}{\wht{a}^{\dagger}}
\newcommand{\down}{\wht{a}}
\renewcommand{\vec}[1]{\boldsymbol{#1}} 
\newcommand{\Tr}[1]{\operatorname{tr}\left( #1 \right)}

\newcommand{\zeile}[1]{\vskip #1 \baselineskip} 

\newcommand{\appsection}[2]{\setcounter{equation}{0}\setcounter{subsection}{0}
\section*{Appendix #1. #2}
\renewcommand{\theequation}{#1.\arabic{equation}}
              \renewcommand{\thesection}{#1} }

\catcode`\@=11
\def\numberbysection{\@addtoreset{equation}{section}
        \def\theequation{\thesection.\arabic{equation}}}
\numberbysection

\begin{document}

\begin{titlepage}

\vskip 1.5 cm
\begin{center}
{\Large \bf Lindblad dynamics of a quantum spherical spin}
\end{center}

\vskip 2.0 cm
\centerline{{\bf Sascha Wald} and {\bf Malte Henkel}}
\vskip 0.5 cm
\begin{center}
Groupe de Physique Statistique, \\
D\'epartement de Physique de la Mati\`ere et des Mat\'eriaux, \\
Institut Jean Lamour (CNRS UMR 7198),  Universit\'e de Lorraine Nancy, \\ 
B.P. 70239, F -- 54506 Vand{\oe}uvre l\`es Nancy Cedex, France
\end{center}

\begin{abstract}
The coherent quantum dynamics of a single bosonic spin variable, subject to a constraint 
derived from the quantum spherical model of a ferromagnet,
and coupled to an external heat bath, is studied through the Lindblad equation for the 
reduced density matrix. Closed systems of equations of motion
for several quantum observables are derived and solved exactly. The relationship to the 
single-mode Dicke model from quantum optics is discussed. 
The analysis of the interplay of the quantum fluctuation and the dissipation and their
influence on the relaxation of the time-dependent magnetisation leads to the distinction 
of qualitatively different regimes of weak and strong quantum
couplings. Considering the model's behaviour in an external field as a simple mean-field approximation 
of the dynamics of a quantum spherical ferromagnet, the magnetic
phase diagram appears to be re-entrant and 
presents a quantum analogue of well-established classical examples of fluctuation-induced order. 
\end{abstract}

\vfill
~\\

\end{titlepage}

\section{Introduction}

The statistical mechanics of phase transitions continues to raise many physical and mathematical challenges for
the improved understanding of collective effects, as they arise in systems made up from a large number of strongly
interacting degrees of freedom, either at thermodynamic equilibrium \cite{Amit84,Card96,Sach99,Nish11,Wipf13} 
or else in dynamics  \cite{Card96,Cugl03,Henk09,Henk10,Taeu14}. 
Besides general schemes, such as the renormalisation group or conformal invariance, 
much useful information has been gleaned from the study of exactly solvable models. In this respect, one of the
most-studied models is the so-called {\em `spherical model'} of a ferromagnet \cite{Berl52,Lewi52}. 
It combines the attractive features of 
(i) admitting an exact solution for any space dimension $d$, yet  
(ii) the resulting critical behaviour is generically not of a mean-field type. 
While it is well-established that the original formulation in terms of classical `spherical'
spins $S_i\in\mathbb{R}$ is physically not entirely satisfactory in the limit of low temperatures $T\to 0$, this
difficulty can be eliminated by going over to a quantum spin formulation \cite{Ober72}, which does not modify the
critical behaviour. There exists many detailed studies of the critical behaviour of both the classical and the quantum
version of the spherical model, see e.g. \cite{Joyc72,Henk84a,Niew95,Vojt96,Bran00,Oliv06,Gome13,Wald15}
and references therein.\footnote{For quantitative applications of the spherical model to spin liquids, see \cite{Gara99,Isak04,Berg07,Isak15}.}
Similarly, the slow long-time non-equilibrium dynamics of the classical spherical model has received a lot
of attention, both in studies of glassy \cite{Cugl95,Cugl03,Cham06} 
and non-glassy magnetic systems \cite{Ronc78,Godr00b,Godr13}. 
Recent developments include an analysis of the distribution of global fluctuations \cite{Fort12}, and 
relationships with the distribution of the gaps of the eigenvalues 
of random matrices \cite{Fydo15} and the kinetics of interface growth \cite{Henk15}. 

In this work, we shall consider the dynamics of the {\em quantum} spherical model. 
Since the system is coupled at least to
an external thermal reservoir, a description convenient for an {\em open} quantum system is required. 
Specifically, how to write down a `quantum version' of the Langevin equation of the spherical model~? While in the
statistical mechanics community, such attempts are often considered {\it `\ldots very much model-dependent and difficult
to generalise'} \cite[p. 394]{Cugl03}, these are routinely studied by the quantum optics or mathematics communities, 
see e.g. \cite{Carm99,Breu02,Engl02,Scha14,Atta06,Atta07,Elou15,vanH15} and references therein. 
In these studies, the classical master equation is replaced by the
{\em Lindblad equation}, where the evolution of the time-dependent reduced density matrix of the system is described by a quantum
Liouville operator involving the system's quantum Hamiltonian 
and additional terms which describe the coupling to the bath. 

Before we shall turn to this, let us briefly recall, following \cite{Carm99} 
and generalising to a large number of degrees of freedom, why a straightforward-looking extension 
of a classical Langevin equation is insufficient for the description of coherent quantum dynamics. 
Consider a pre-quantum spherical model, where the dynamical variables are the spherical spin-operators ${S}_{\vec{n}}$ 
(at each site $\vec{n}\in \mathscr{L}$ of a hyper-cubic lattice 
$\mathscr{L}\subset \mathbb{Z}^d$, with ${\cal N}=|\mathscr{L}|$ sites) 
and the canonically conjugate momenta ${P}_{\vec{n}}$. By analogy with \cite{Ober72}, the Hamiltonian is
\BEQ \label{1.1}
H = \sum_{\vec{n}\in\mathscr{L}}\left(\frac{g}{2} P_{\vec{n}}^2+\frac{\mu}{2} S_{\vec{n}}^2 
- \sum_{j=1}^d J S_{\vec{n}} S_{\vec{n}+\vec{e}_j}\right) 
\EEQ
Herein, $g$ is a coupling constant, $\mu$ denotes a Lagrange multiplier, 
to be found self-consistently from the spherical constraint
$\left<\sum_{\vec{n}}{S}_{\vec{n}}^2 \right> = {\cal N}$, and $\vec{e}_j$ is the $j^{\rm th}$ cartesian unit vector. 
Taking into account the momenta, one may write down a Kramers equation \cite{Tail06} 
\BEQ \label{1.2}
\partial_t {S}_{\vec{n}} =  \left\{ {S}_{\vec{n}} , H \right\} \;\; , \;\;  
\partial_t {P}_{\vec{n}} =  \left\{ {P}_{\vec{n}} , H \right\} -\gamma {P}_{\vec{n}} + {\eta}_{\vec{n}}(t)
\EEQ
with a damping parameter\footnote{Throughout, units are such that the  Boltzmann constant $k_{\rm B}=1$.}   
$\gamma$ and the standard centred white-noise $\eta_{\vec{n}}(t)$, with correlator 
$\left<\eta_{\vec{n}}(t)\eta_{\vec{m}}(t')\right> =2\gamma T \delta_{\vec{nm}}\delta(t-t')$. 
Herein, the brackets $\{.,.\}$ denote the Poisson brackets. 
Eq.~(\ref{1.2}) is a well-defined and interesting dynamics with non-trivial properties, 
such as a hidden super-symmetry \cite{Tail06}. 
It might also appear as a natural starting point for going over to the dynamics of the quantum case. 
According to the natural-looking procedure suggested in \cite{Shuk81}, one replaces in (\ref{1.2}) 
(i) the classical variables $S_{\vec{n}}\mapsto \wht{s}_{\vec{n}}(t)$ and $P_{\vec{n}}\mapsto \wht{p}_{\vec{n}}(t)$ by
time-dependent operators $\wht{s}_{\vec{n}}(t)$ and $\wht{p}_{\vec{n}}(t)$, which 
(ii) at the initial time $t=0$ obey the canonical equal-time commutation relations 
$\left[\wht{s}_{\vec{n}}(0), \wht{p}_{\vec{m}}(0)\right]=\II\hbar \delta_{\vec{n},\vec{m}}$,  
(iii) replaces the Poisson brackets $\{.,.\} \mapsto \frac{1}{\II\hbar}\left[.,.\right]$ by a commutator and
(iv) introduces a noise operator $\wht{\eta}_{\vec{n}}(t)= \eta_{\vec{n}}(t)\wht{\bf 1}$. Applied to the
spherical model Hamiltonian (\ref{1.1}), this procedure would lead 
to the quantum operator equations of motion \cite{Shuk81}
\BEA
\partial_t  \wht{s}_{\vec{n}} &=& g \wht{p}_{\vec{n}} \nonumber \\
\partial_t  \wht{p}_{\vec{n}} &=& - \mu \wht{s}_{\vec{n}} - \gamma g  \wht{p}_{n}
+\frac{J}{\mu}\sum_{j=1}^d \left( \wht{s}_{\vec{n}-\vec{e}_j} 
+ \wht{s}_{\vec{n}+\vec{e}_j}\right) +  \wht{\eta}_{\vec{n}} 
\label{1.3}
\EEA
However, if one defines the commutator 
$\wht{c}_{\vec{n}}(t) := \left[ \wht{s}_{\vec{n}}(t), \wht{p}_{\vec{n}}(t)\right]$, 
one promptly obtains the equation of motion 
\BEQ
\partial_t \wht{c}_{\vec{n}}(t) = -g \gamma\: \wht{c}_{\vec{n}}(t)
\EEQ
Hence, $\wht{c}_{\vec{n}}(t) = \II\hbar\; e^{-t/t_{\rm deco}}$ which means that the dynamics (\ref{1.3}) 
dissipates the quantum structure on a finite time-scale, of order $t_{\rm deco}=1/(\gamma g)$ \cite{Carm99}.\footnote{Formally, 
one might introduce an `effective Planck constant' $\hbar_{\rm eff}(t) = \hbar e^{-t/t_{\rm deco}}$ decaying to zero.} 
Indeed, one may define more general quantum spherical models 
by adding interactions between the momenta into the Hamiltonian
(\ref{1.1}), see eq.~(\ref{A1}) in appendix~A. At equilibrium, this is known to lead to new quantum effects, 
such as a re-entrant quantum phase transitions in sufficiently small
dimensions $d\lesssim 2.065$ \cite{Wald15}. However, as we shall show in appendix~A, 
a corresponding generalisation of the 
equations of motion (\ref{1.3}) always leads, for times $t\gg t_{\rm deco}$, 
back to the well-known relaxational dynamics \cite{Ronc78,Godr00b,Dura15} 
of the classical spherical model with $g=0$. In consequence, any quantum effects of the equilibrium state are washed out
by this incoherent dynamics, which does not even relax to the required equilibrium state. 

If a dynamical description is sought which maintains the quantum coherence 
of an open quantum system with $\gamma>0$ and $g>0$, and evolves towards the correct quantum equilibrium state, 
a different approach is required. Here, we shall adopt the result 
of a profound analysis of the interactions of the system with its
environment, see e.g. \cite{Carm99,Breu02,Engl02,Atta06,Atta07,Scha14} and references 
therein, and shall take as our starting point 
the {\em Lindblad equation} for the time-dependent density matrix $\wht{\rho}=\wht{\rho}(t)$ of the system
\BEQ \label{1.5}
\frac{\D\wht{\rho}}{\D t} = -\frac{\II}{\hbar} \left[ \wht{H}, \wht{\rho} \right] 
- \sum_{\alpha} \left( L_{\alpha} \wht{\rho} L_{\alpha}^{\dag} 
-\demi L_{\alpha}^{\dag} L_{\alpha} \wht{\rho} -\demi \wht{\rho} L_{\alpha}^{\dag} L_{\alpha} \right)
\EEQ
where the Lindblad operators $L_{\alpha}$ describe the damping through 
the coupling of the system to the reservoir. It is
well-known that the Lindblad equation preserves the trace, the hermiticity 
and the positivity of the density matrix \cite{Carm99,Breu02,Scha14}. 
In section~2, we shall specify the Lindblad operators completely and
shall also re-derive that the Lindblad equation 
dynamics preserves the canonical commutator relations between spins and momenta, at least on average.\footnote{From the
point of view of classical dynamics, one might say that the Lindblad equation (\ref{1.5}) automatically contains a large number of
conserved quantities, corresponding to the canonical commutators.}  

In this work, we shall consider a thermal reservoir of bosonic particles and the $L_{\alpha}$ will be chosen
accordingly (see section~2). In the past, much work has been done on systems with only two energy levels per site. 
Remarkably, for several quantum chains, exact results on the non-equilibrium stationary states have been derived 
through techniques of quantum integrability \cite{Pros11,Kare13,Popk13,Braa13,Batc15}, see \cite{Pros15,Braa15} for recent reviews. 
Here, we present first results of an exploration of a quantum system where the space of states of a single site is larger. 
Indeed, we hope to make use of the solvability of the quantum spherical model in order to construct 
exact solutions of the corresponding Lindblad equation. 
The aim of such an approach should be a comparative analysis of quantum 
vs classical phase transitions, see \cite{Podo14} for
an example in the Ising model universality class. 
As a first step towards the realisation of this programme, 
we consider here the quantum dynamics of a {\em single} spherical spin,
which might also be viewed as a simple mean-field solution of the dynamics of a $N$-body problem and to 
discuss the resulting phase diagram in the stationary state. Surprisingly, it appears that the resulting  
phase-diagram appears to be re-entrant and thereby presents a quantum analogue of a mechanism, 
well-known from classical systems \cite{Katz84,Helb00,Zia02,Radz03,Ehre10,Borc14}, 
where it is often referred to as `{\em freezing-by-heating}' or `{\em getting more from pushing less}'. 

This work is organised as follows. In section~2, we precisely define the single-spin quantum spherical model 
and derive from (\ref{1.5}) the quantum equations of motion of several observables. We also comment on its relationship
with the Dicke model. 
In section~3, the equations of motion are solved exactly at temperature $T=0$ and
for a vanishing external field. 
In section~4, an external field is included and this is used to derive
a quantum mean-field theory of the non-equilibrium stationary state. 
The r\^ole of the couplings $g$ and $\gamma$ on the phase diagram, 
as well as the corrections implied by a sufficiently small temperature $T>0$ will be discussed. 
Section~5 gives our conclusions. 
Several appendices treat technical details of the calculations. 

\section{The Model}

\subsection{A single spherical quantum spin}

The Hamiltonian of a single spherical quantum spin ({\sc sqs}), in an external magnetic field $B$, reads \cite{Ober72}
\BEQ \label{2.1}
\wht{H} = \frac{g}{2} \hp^{\,2} + \frac{\mu}{2}\hs^{\,2} -B\hs
\EEQ
with the canonical commutation relation $\left[ \hs,\hp\, \right] = \II\hbar$. This is the quantum version of
the classical Hamiltonian (\ref{1.1}), reduced to a single degree of freedom. 
Herein, $g$ is the quantum coupling of the system with the classical limit $g\rightarrow0$. 
The \emph{Lagrange multiplier} $\mu=\mu(t)$ is chosen\footnote{The time-dependence of $\mu(t)$ is essential for a 
correct description of the relaxation properties \cite{Ronc78,Cugl95,Godr00b}, in contrast to the approach followed in \cite{Shuk81}.} 
to ensure the (mean) spherical constraint \cite{Lewi52}
\BEQ\label{eq:sc}
\left< \wht{s}^{\,2}\right> = 1 \ .
\EEQ
Consequently, $\mu/2 \cdot \hs^{\,2}$ is simply an effective energy shift of the Hamiltonian. 

It is convenient to go over to creation and annihilation operators, in the usual way \cite{Ober72}
\BEQ
\hs = \sqrt{\frac{\hbar g}{2\omega}\,}\left(\up+\down\right) \;\; , \;\; 
\hp = \II \sqrt{\frac{\hbar \omega }{2 g}\,}\left(\up-\down\right), \hspace{.5cm} \textrm{with}
\hspace{.5cm}
\omega = \omega(t) := \sqrt{\mu(t)g\,}
\label{2.3}
\EEQ
which recasts the Hamiltonian into the form 
\BEQ \label{2.4}
\wht{H} = \hbar \omega (t) \left(\up\down +\demi \right) 
- B \sqrt{\frac{\hbar g}{2\omega}\,}\,\left(\up+\down\right) \ .
\EEQ
The spherical constraint (\ref{eq:sc})  
introduces a functional relationship between the (effectively time-dependent) frequency 
and the two-particle-operator expectation values, via
\BEQ\label{eq:constraint}
\omega =\omega(t) = \frac{\hbar g }{2} \left(\left< \up\up\right> + 
\left< \down\down\right> +2\left< \up\down\right> + \stackrel{~}{\stackrel{}{1}}\right)
\EEQ
This condition, along with the explicit Hamiltonian (\ref{2.4}), defines the closed system completely.

If coupled to an external bath, a coherent quantum dynamics is formulated by adopting a Schr\"odinger picture and 
writing down a Lindblad equation for the time-dependent
density matrix $\wht{\rho}=\wht{\rho}(t)$ of this open quantum system. We assume
a thermal coupling to the zero-field modes \cite{Carm99,Breu02,Scha14} and consider the following
Lindblad equation 
\BEQ \label{Gl:Lindblad}
\dot{\dicht} = -\frac{\II}{\hbar}\left[\wht{H}, \dicht\right] + \gamma (n_{\omega}+1)\left[ \down \dicht \up
-\demi \left(\up\down\,\dicht+\dicht\,\up\,\down\,\right)\right] +\gamma n_{\omega}  \left[
\up\dicht \down
-\demi \left(\down\up\,\dicht+\dicht\,\down\,\up\right)\right]
\EEQ
where the bath, of given temperature $T$, is characterised by the Bose-Einstein statistics 
$n_{\omega} = \left( e^{\hbar\omega/T}-1\right)^{-1}$ and $\gamma$ is a coupling constant. 
Because of the spherical constraint (\ref{eq:constraint}), the frequency $\omega=\omega(t)$ must be considered as
a time-dependent function. In consequence, 
the occupation number $n_{\omega}=n_{\omega(t)}$ becomes effectively time-dependent as well.

The three equations (\ref{2.4},\ref{eq:constraint},\ref{Gl:Lindblad}) 
define completely our time-dependent, open, quantum model system, of a single {\sc sqs}. 
It depends on the physical parameters temperature $T$, magnetic field $B$, dissipation coupling $\gamma$ and
quantum coupling $g$. We shall consider these equations as a phenomenological 
ansatz and shall concentrate from now on how to extract their
time-dependent behaviour.

Consequently, we deduce the closed set of equations of motion for the following averages 
\BEA \label{eq:aa}
\partial_t\left<\down\down\,\right> &=& -2\left[ \frac{\gamma}{2} + \II \omega \right]\left<\down\down\,\right> 
+\II\sqrt{\frac{2g}{\hbar\omega}\,}\,B \left< \down\, \right> \\
\label{eq:adaggera}
\partial_t\left<\up\down\right> &=& - \gamma \left<\up\down\right> +\gamma n_{\omega}
+\II \sqrt{\frac{g}{2\hbar\omega}\,}\,{B}\left(\left< \up \right>- \left< \down\, \right>\right)
\\
\label{eq:a}
\partial_t \left<\down\,\right>  &=& -\left[ \frac{\gamma}{2} + \II \omega\right]\left<\down\,\right> 
+\II \sqrt{\frac{g}{2\hbar\omega}\,}\,{B}
\EEA
where $\omega=\omega(t)$ is given by (\ref{eq:constraint}). 
Clearly, $\left<\up\right> = \left<\down\,\right>^*$ and $\left<\up\up\right>=\left< \down\down\,\right>^*$. 

In the chosen form (\ref{Gl:Lindblad}) of the Lindblad equation, 
where the bosonic creation and annihilation operators $\up$ and $\down$ are guaranteed to be
time-independent, we can now briefly comment on the preservation of the quantum coherence. 
Specifically, the 
average of their commutator becomes
\BEQ
\left<\left[\down,\up \right] \right> (t) = \Tr{\left[\down,\up \right]\dicht(t)} 
= \Tr{\wht{\bf 1}\dicht(t)} = 1
\EEQ
Inverting (\ref{2.3}), it also follows that the canonical commutation relations between 
$\hs$ and $\hp$ are kept, at least on average, 
viz. $\left< \left[ \hs, \hp \right]\right> = \II \hbar$ for all times $t$. 

Conceptually, the constraint (\ref{eq:sc}) means that besides the coupling to an external thermal bath with a fixed 
temperature $T$, as described by the dissipative terms in (\ref{Gl:Lindblad}), 
effectively there is a second external bath which acts on the system in a way that (\ref{eq:sc}) holds
true, where $\mu$ is canonically conjugate to $\hs^{\,2}$. 
In principle, we could have followed the standard derivation of Lindblad equations, see \cite{Carm99,Breu02,Atta07,Scha14}, 
in order to obtain explicitly the corresponding Lindblad operators $L_{\alpha}$, as in eq.~(\ref{1.5}). 
We shall not carry this out here, since we expect that for a large number of degrees of freedom, this explicit construction
would merely correspond to a change of the statistical ensemble. 
That should be analogous to a change between, say, canonical and
grand canonical ensembles. In the classical spherical model, this would correspond to requiring the spherical constraint
either exactly on each microscopic spin  configuration, or merely on average. 
It is well-known that this distinction becomes unimportant for the analysis of the critical behaviour in the
limit ${\cal N}\to\infty$ of a large number of spins, both at and far from equilibrium 
\cite{Lewi52,Fusc02}.\footnote{Since we consider this study as preliminary work on the dynamics of the 
quantum spherical model with ${\cal N}\to\infty$ spins, we do not go further into the distinction of ensembles. 
In this respect, the present results on a single spherical spin should rather be viewed as some mean-field approximation
of that full ${\cal N}$-body problem.}

Turning now to the analysis of the long-time behaviour following from the equations of motion (\ref{eq:aa},\ref{eq:adaggera},\ref{eq:a}),
we keep in mind that the combined action of two distinct external baths may lead the system to evolve 
towards a non-equilibrium stationary state. 

\subsection{Relationship with the Dicke model}

The {\em single-mode Dicke model} \cite{Dick54} describes the cooperative interaction of $\cal M$ atoms in a cavity with a
single mode of the radiation field. In the rotating-wave-approximation ({\sc rpa}), the Dicke model 
Hamiltonian reads  \cite{Garr11,Ton09} (in this section, we set $\hbar=1$ throughout)
\BEQ \label{2.11}
\wht{H}_D = \omega\left( \wht{S}_z + \frac{{\cal M}+1}{2} \right) 
+ \frac{\gamma}{\sqrt{\cal M}\,}\left( \wht{r}^{\, \dagger}\wht{S}_- + \wht{r} \wht{S}_+  \right) + \omega_r \wht{r}^{\,\dagger} \wht{r} 
\EEQ
Herein, the $\cal M$ atoms are each represented by a two-level system, with ground states $\ket{g_j}$ and excited states
$\ket{e_j}$, $j=1,\ldots,{\cal M}$. The transitions in each atom are described in terms of spin-$\demi$ operators 
\BEQ
\wht{S}^{(j)}_{+} = \ket{e_j}\bra{g_j} \;\; , \;\; 
\wht{S}^{(j)}_{-} = \ket{g_j}\bra{e_j} \;\; , \;\;   
\wht{S}^{(j)}_{z} = \frac{1}{2}\left( \ket{e_j}\bra{e_j}  - \ket{g_j}\bra{g_j} \right)
\EEQ
and the collective atomic operators read $\wht{S}_{\pm} = \sum_{j=1}^{\mathcal{M}} \wht{S}_{\pm}^{(j)}$ and 
$\wht{S}_z = \sum_{j=1}^{\mathcal{M}} \wht{S}_{z}^{(j)}$. The state of the cavity (reservoir) is described by 
the bosonic raising and lowering operators $\wht{r}^{\, \dagger}$ and $\wht{r}$, with 
$\left[ \wht{r},\wht{r}^{\,\dagger}\right]=1$. The two energy scales $\omega$ and $\omega_r$, as well as the coupling $\gamma$, are taken to be constants. 
The Dicke model is known to undergo a continuous phase transition, from a normal to a super-radiant phase, with the order parameter
$\lim_{{\cal M}\to \infty} {\cal M}^{-1}\left\langle \wht{r}^{\,\dagger}\wht{r} \right\rangle$. This transition either occurs
at a finite critical temperature $T_c>0$ and then is thermally  driven, or else is a quantum phase transition at $T=0$ and is driven
by $\gamma$, see the reviews \cite{Garr11,Ton09} and references therein.

Within this setup, the model can be re-written through a Holstein-Primakoff transformation
\cite{Hol40,Garr11}, which replaces the collective atomic operators by a bosonic degree of freedom
via
\BEQ \label{2.13}
\wht{S}_{+} = \up \mathcal{M}^{1/2}\left(1-\up\down/\mathcal{M} \right)^{1/2} \;\;,\;\; 
\wht{S}_{-} = \mathcal{M}^{1/2}\left(1-\up\down/\mathcal{M} \right)^{1/2}\down \;\;,\;\;
\wht{S}_{z}  = \up\down  - \mathcal{M}/2
\EEQ
where we recognise the system's bosonic creation and annihilation operators $\up$ and $\down$. 

Inserting (\ref{2.13}) into (\ref{2.11}) and expanding this up to leading order in $\up\down/\mathcal{M}$, 
gives in the limit ${\cal M}\to\infty$ the effective low-energy Hamiltonian 
\BEQ \label{2.14}
\wht{H}_D \approx  \omega \left( \up\down +\demi\right)
 + \gamma\left( \wht{r}^{\, \dagger} \down  + \wht{r}\up \right) + \omega_r \wht{r}^{\, \dagger} \wht{r} 
\EEQ
This Hamiltonian has the general form $\wht{H}_D = \wht{H}_{\rm sys} + \wht{H}_{\rm int} + \wht{H}_{\rm res}$ and describes, 
as a `system', a single boson with Hamiltonian $\wht{H}_{\rm sys}$ analogous to (\ref{2.4}), 
interacting through the term $\wht{H}_{\rm int}$ with a `\textit{bosonic single-mode reservoir}' 
described itself by $\wht{H}_{\rm res}$. 
This is the usual starting point for deriving a
Lindblad equation for the dynamics of the `system'  by tracing out the degrees of freedom of the `reservoir'.
Indeed, if one adopts the usual procedure of fixing the properties of the `reservoir', 
for instance its temperature $T$, and also assumes the `reservoir' large enough as to be not influenced by the properties of the `system', 
a lengthy but standard calculation shows that the quantum dynamics of the reduced density matrix of the
`system' is given by the Lindblad equation  (\ref{Gl:Lindblad}) with the Hamiltonian (\ref{2.4}), with $B=0$
\cite{Breu02,Carm99,Scha14}. 

In spite of this formal analogy, the {\sc sqs} and the single-mode Dicke model are still different. 
First, the phase transition in the Dicke model refers for the definition of the order parameter to
the properties of the `reservoir', which is traced out in the {\sc sqs}. This is probably not very important, since in
the low-energy Hamiltonian (\ref{2.14}), `system' and `reservoir' can be exchanged 
according to $(\down,\up)\longleftrightarrow (\wht{r},\wht{r}^{\,\dagger})$, 
along with $\omega \longleftrightarrow \omega_r$. In this respect, the {\sc sqs} and the Dicke model are dual to each other. 
Second, this averaging is normally done for a fixed temperature and the assumed properties of 
the reservoir in general do not
allow for a phase transition.\footnote{We hope to return elsewhere to an exploration of the consequences of
an assumed phase transition in an external reservoir.} 
Third, and probably most important, the Dicke model considers the angular frequency $\omega$ 
as a fixed parameter, whereas in the {\sc sqs}, its time-dependent value $\omega(t)$ has to
found self-consistently from the spherical constraint (\ref{eq:constraint}). In the next section, we shall
see that this leads to an important qualitative difference in the behaviour of the two models.\footnote{In the limit $\omega_r\to 0$, the
Dicke Hamiltonian (\ref{2.11}) becomes the one of the integrable Tavis-Cummings model, whose dynamics for an arbitrarily prescribed time-dependent
$\omega=\omega(t)$ can be studied through the Bethe ansatz \cite{Barm13}.}

\section{Analytic solution in zero field and at zero temperature}

We focus on the case where $B=0$ and $T=0$.
This particular case is analytically solvable for all times $t$.

Due to the vanishing field $B=0$, the equations (\ref{eq:aa},\ref{eq:adaggera}) decouple from (\ref{eq:a}), such that
the single-particle operators can be treated separately. We must investigate the system
\BEA
\partial_t\left<\down\down\,\right> &=& -\left[ \gamma +2\II \omega (t) \right]\left<\down\down\,\right> \\
\partial_t\left<\up\down\,\right> &=& - \gamma \left<\up\down\,\right> \ .
\EEA
Obviously, the particle-number-operator expectation value decays exponentially
\BEQ \label{3.3}
\left< \up\down\,\right> =N e^{-\gamma t}\ , \hspace{.5cm} N\in \mathbb{R}_+ \ .
\EEQ
By means of this solution and the spherical constraint (\ref{eq:constraint}), 
the expectation value of the pair-annihilation operator obeys the equation
\BEQ
\partial_t\left< \down\down\, \right> = 
-\left[\gamma + \II\hbar\left(1+2Ne^{-\gamma t}\right) \right]\left< \down\down\, \right> 
-\II\hbar g \left|\left< \down\down\, \right> \right|^2
-\II\hbar g \left< \down\down\, \right>^2 \ .
\EEQ
Separating amplitude and complex phase, via $\left< \down\down\, \right> = R(t)e^{\II\Theta(t)}$, leads to
\BEA
\label{eq:amplitude}
\dot{R}(t) &=&-\gamma\: R(t) \\ 
\label{eq:phase}
\dot{\Theta}(t) &=& -\hbar g \left[ 2R(t)\cos\Theta(t) + 2N\,e^{-\gamma t} +1\right] \ .
\EEA
These equations allow to separate the two basic physical mechanisms and show in particular 
that the exponential decay is an intrinsic fact of the \textit{classical spherical model}, whereas the time-dependent 
phase $\Theta(t)$ is a quantum effect of the {\sc sqs} (for $\hbar g=0$, eq. (\ref{eq:phase}) simply states that $\dot{\Theta} = 0$).

The amplitude equation (\ref{eq:amplitude}) simply gives $R=R(t) = Ae^{-\gamma t}$, 
with $A\in \mathbb{R}_+$.\footnote{At equilibrium, it follows from the Hamiltonian (\ref{2.1}) that
$\left\langle\down \down\,\right\rangle_{\rm eq}=\left\langle\up\up\right\rangle_{\rm eq}=0$. 
Hence the amplitude $A$ can be viewed as a measure of the initial distance from the equilibrium state.} 
However, the phase equation is more complicated. In appendix~B, we show that the solution of (\ref{eq:phase}) is 
\BEQ \label{eq:cosTneq}
\cos\Theta = \Re\hspace{-.2cm}\left(-\frac{N}{A}-\II\sqrt{1-\frac{N^2}{A^2}}
+\frac{\hbar g}{\gamma} \left(\sqrt{1-\frac{N^2}{A^2}}-\II\frac{N}{A}\right)
\frac{\II\frac{\gamma}{\hbar g} K M(\mathscr{T}_{(1,1)})-U(\mathscr{T}_{(1,1)})}{KM(\mathscr{T})-U(\mathscr{T})}\right)
\EEQ
for $A\neq N $ and 
\BEQ \label{eq:cosTeq}
\cos\Theta = -\Re\left( 1 + \frac{\II}{\sqrt{A}}\: \e^{\frac{\gamma}{2} t}
\frac{K J_{\II\frac{\hbar g}{\gamma}}\left(2\II\frac{\hbar g}{\gamma}\sqrt{Ae^{-\gamma t}}\right)
-J_{-\II\frac{\hbar g}{\gamma}}\left(2\II\frac{\hbar g}{\gamma}\sqrt{Ae^{-\gamma t}}\right)}
{K J_{1+\II\frac{\hbar g}{\gamma}}\left(2\II\frac{\hbar g}{\gamma}\sqrt{Ae^{-\gamma t}}\right)
+J_{-1-\II\frac{\hbar g}{\gamma}}\left(2\II\frac{\hbar g}{\gamma}\sqrt{Ae^{-\gamma t}}\right)}\right)
\EEQ
for $A = N$, respectively.
Herein, the constant $K$ 
is related to the initial condition, $M=M(\mathscr{T})$, $U=U(\mathscr{T})$ 
are Kummer's hypergeometric functions \cite{Abra65}, with the triple argument 
\BEQ 
\mathscr{T} : = \left(-\frac{\hbar g}{2\gamma}\left(\II+\frac{1}{\sqrt{A^2/N^2 -1\,}}\right) \hspace{2mm} ; 
\hspace{2mm}
-\II\frac{\hbar g}{\gamma} \hspace{2mm} ; \hspace{2mm}
2\frac{\hbar g }{\gamma} \sqrt{A^2-N^2\,}\: e^{-\gamma t} \right) 
\EEQ
and the further abbreviation $\mathscr{T}_{(x,y)} := \mathscr{T} + (x;y;0)$. Furthermore $J_p(z)$ denotes the Bessel
function of the first kind and order\footnote{Confluent hypergeometric or Bessel functions with complex indices/orders are often
met in the dynamics of quantum systems, see e.g. \cite{Garr11,Batc15,Braa15}.} $p$. 
We have checked that $-1\leq\cos\Theta(t)\leq 1$ for all
times and all ratios $A/N$. 

The functions (\ref{eq:cosTneq}) and (\ref{eq:cosTeq}) can be analysed in the long-time limit $t\rightarrow \infty$,
respectively $e^{-\gamma t } \rightarrow 0^+$. A Taylor-series expansion in $e^{-\gamma t}$ yields, to leading order  
\BEQ \label{gl:thetainf}
\cos\Theta \simeq \Re \left\{
-\frac{N}{A}-\II\sqrt{1-\frac{N^2}{A^2}}+
\epsilon_1 \cos \hbar g t + \epsilon_2 \sin \hbar g t\right\}
\EEQ
with appropriate (complex) constants $\epsilon_{1}$ and $\epsilon_{2}$. Thus the long-time limit provides the expected
harmonic oscillator with frequency $\Omega = \hbar g$. This asymptotic expansion reveals the oscillations at 
least for large times in the effective frequency $\omega(t)$ (while for all other models like the Dicke model or the Jaynes-Cummings model, the 
frequency $\omega$ is a constant). As the effective oscillation frequency $\omega(t)$ tends to zero for $\hbar g \rightarrow 0$, we observe
here a \textit{quantum effect} of the system.

\begin{figure}[ht!]
 \centerline{\psfig{figure=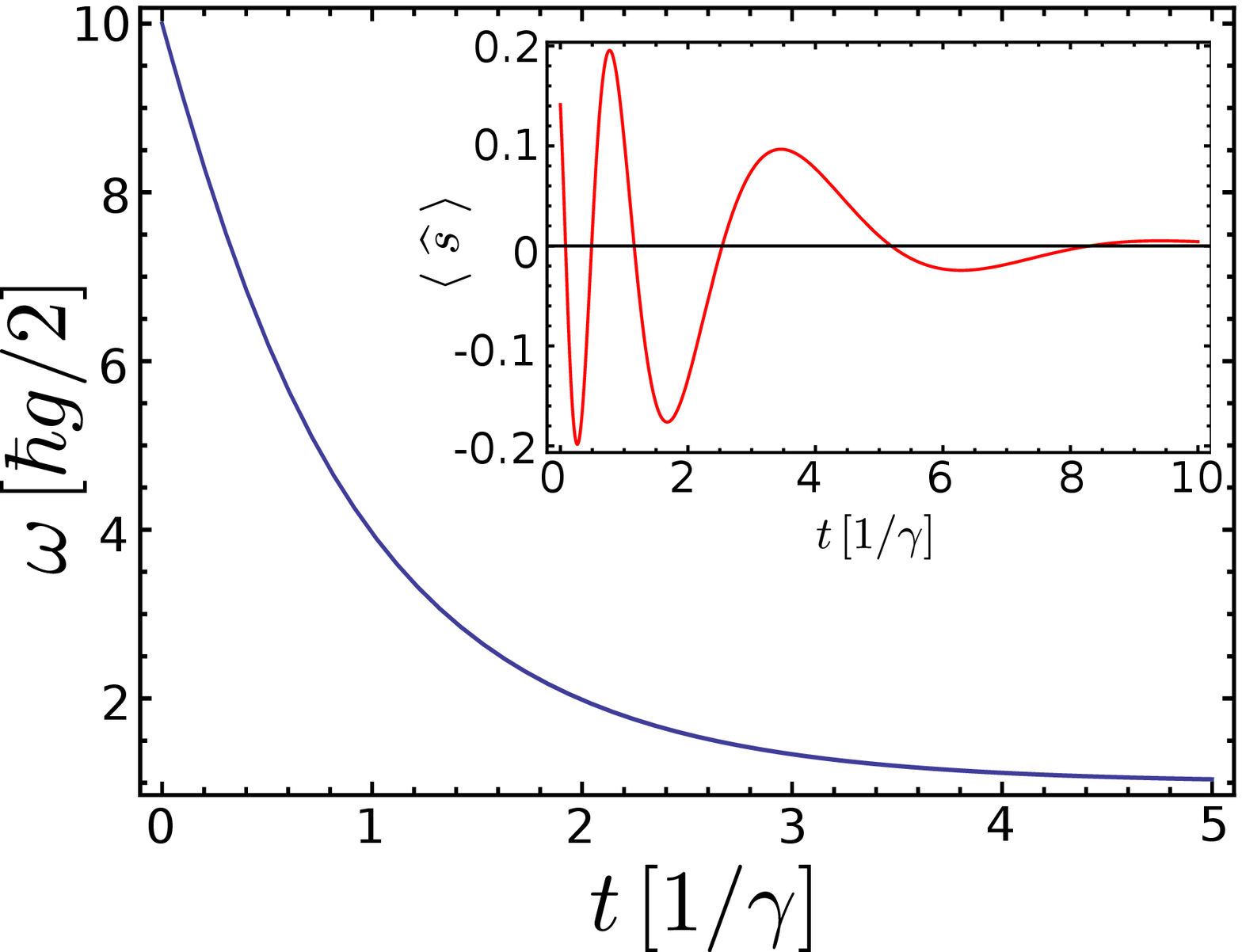,width=3.5in,clip=} ~~\psfig{figure=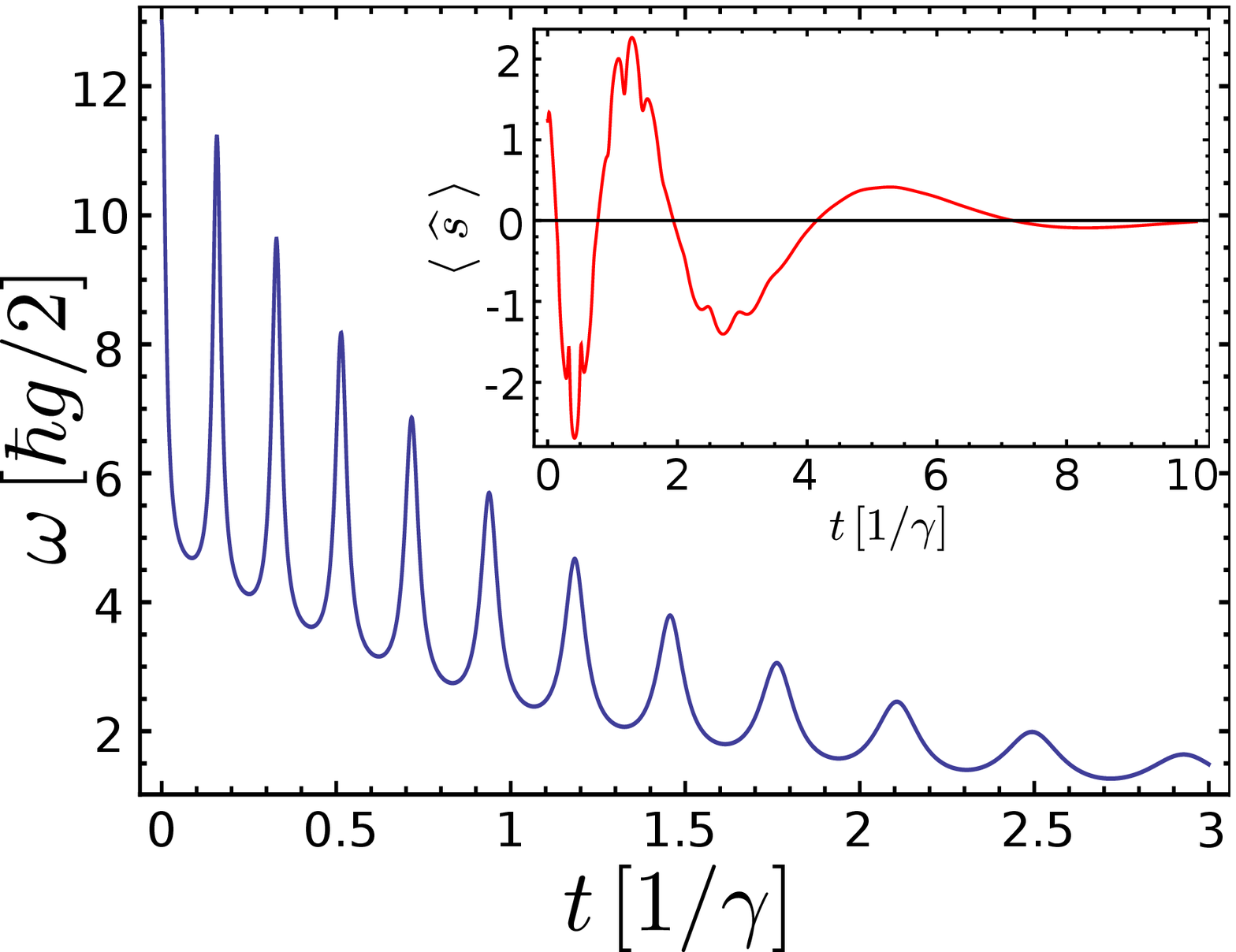,width=3.5in,clip=}}
 \caption[figomega1]{\underline{Left panel:} Time-dependence of the effective frequency $\omega(t)$ (main plot) 
 and of the magnetisation $m(t)=\left< \wht{s}\,\right>(t)$ (inset), for the parameters 
 $A=N=3$ and $g=0.1 J/\hbar^2$ in the weak quantum-coupling regime. 
 A simple exponential decay for the frequency is seen, which leads 
 to a time-varying oscillation frequency in the magnetisation.  
 \underline{Right panel:} analogous plots with parameters $A=N=4$ and $g=10J/\hbar^2$ 
 in the strong quantum-coupling regime. 
 Here, the exponential decay of $\omega(t)$ is modulated by strong oscillations
 with sharp peaks. These lead to a rather complex oscillatory behaviour in the magnetisation.
 }
\label{fig:omega_ana}
\end{figure}

Now, combining eqs.~(\ref{eq:cosTneq}, \ref{eq:cosTeq}) with the definition of 
$\left< \down\down \right>=R(t)e^{\II\Theta(t)}$, the time-dependent
effective frequency $\omega=\omega(t)$ can be reconstructed from eq.~(\ref{eq:constraint}), 
using also (\ref{3.3}). Afterwards, the
magnetisation $m(t)=\left< \wht{s}(t)\right> 
= \sqrt{\frac{\hbar g}{2\omega(t)}\,} \left( \left< \up\right> + \left< \down\right>\right)$,
see (\ref{2.3}), follows by integrating eq.~(\ref{eq:a}).  
Fig.~\ref{fig:omega_ana} shows the resulting oscillation frequency $\omega(t)$ and the magnetisation $m(t)$, 
for the special case $A=N$. 
Already in this more simple case, we observe a distinction between 
(i) a {\em weak-quantum-coupling regime} $g\ll 1 J /\hbar^2$, characterised by a
simple monotonous decay of $\omega(t)$ and a simple oscillatory relaxation of $m(t)$ and 
(ii) a {\em strong-quantum-coupling regime}
$g \gg 1J /\hbar^2$, where on the decay of $\omega(t)$ is superposed strongly peaked non-harmonic oscillations,
which leads to a complex oscillatory behaviour of $m(t)$. 

\begin{figure}
 \centerline{\psfig{figure=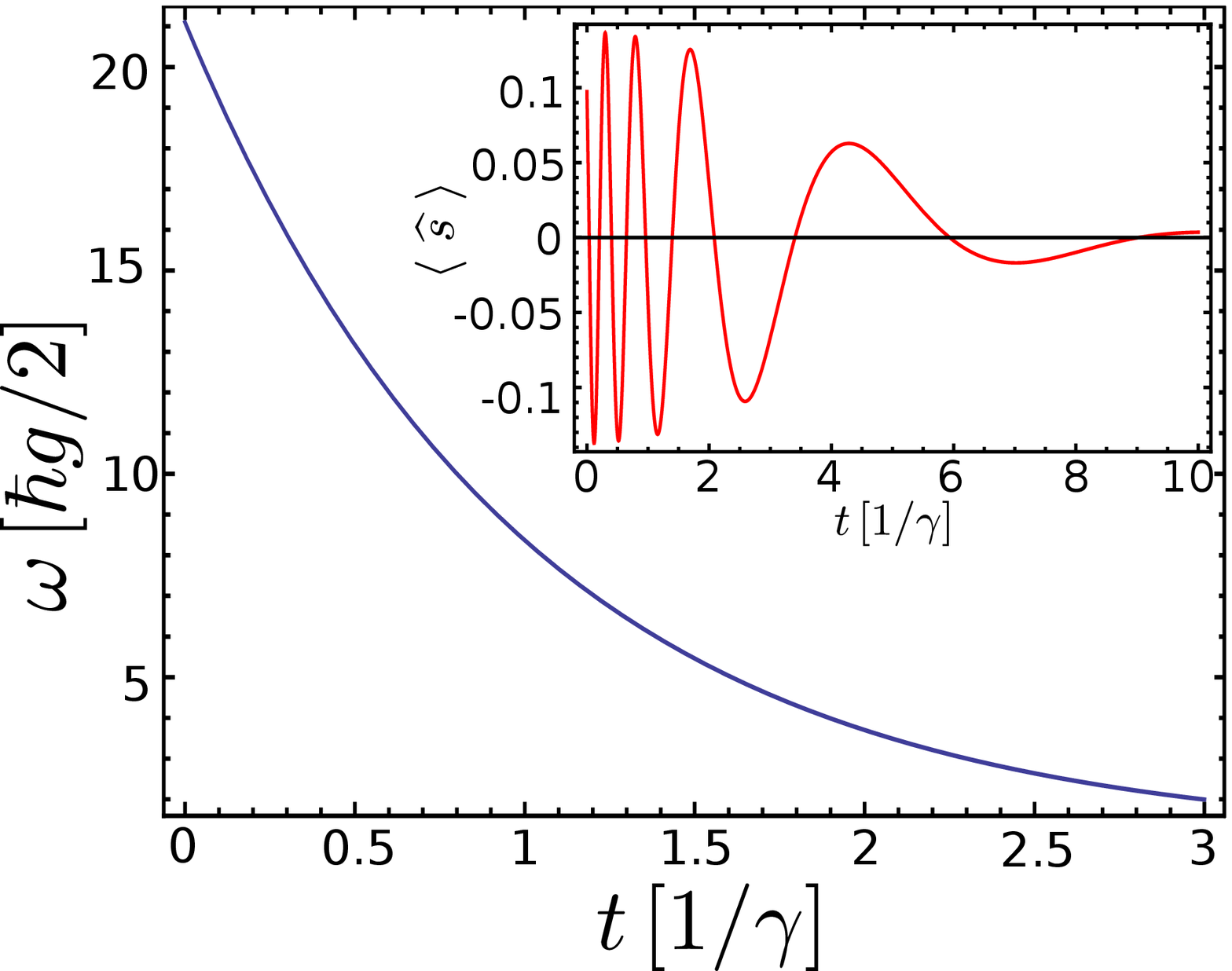,width=2.5in,clip=} ~~\psfig{figure=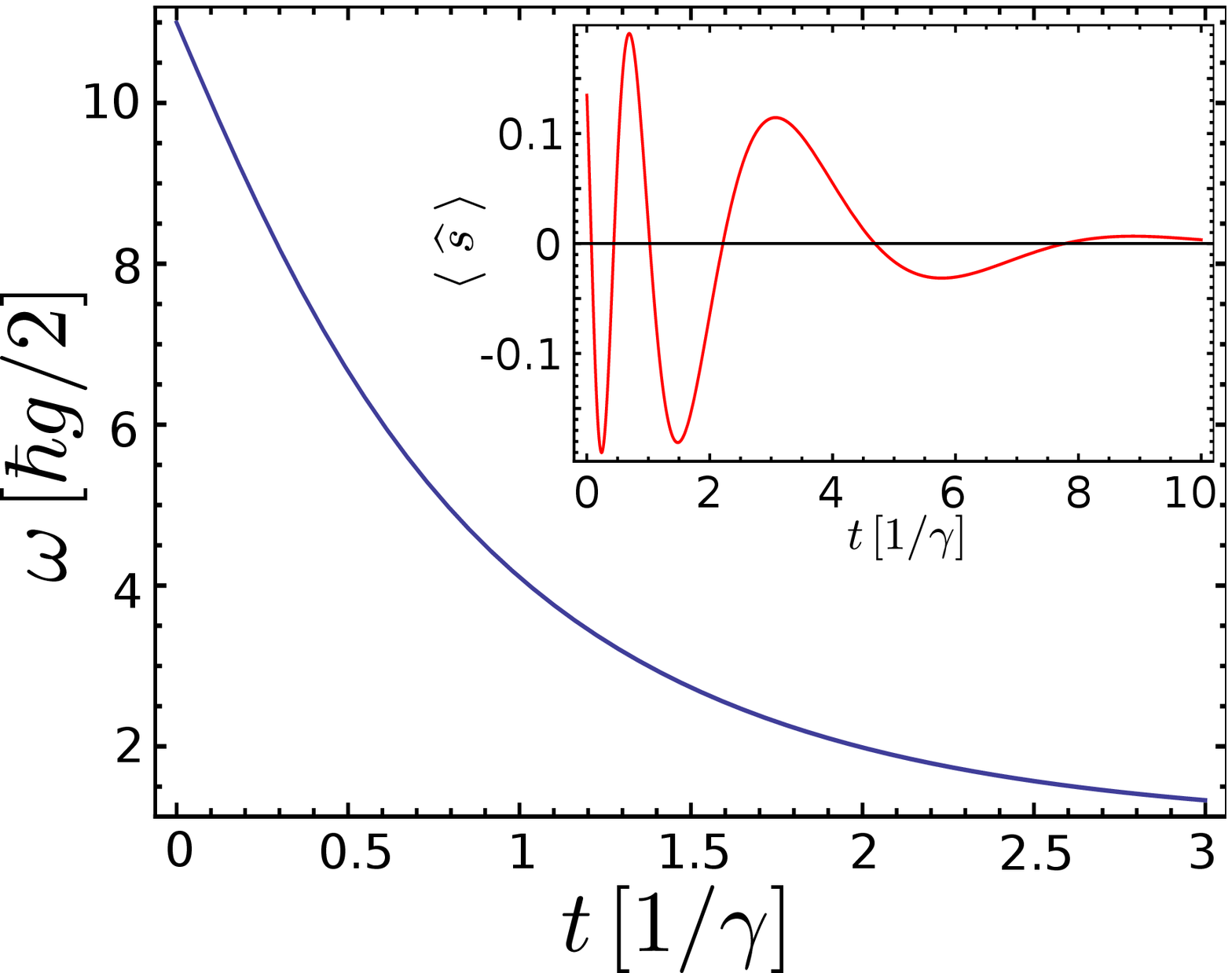,width=2.5in,clip=}
 \psfig{figure=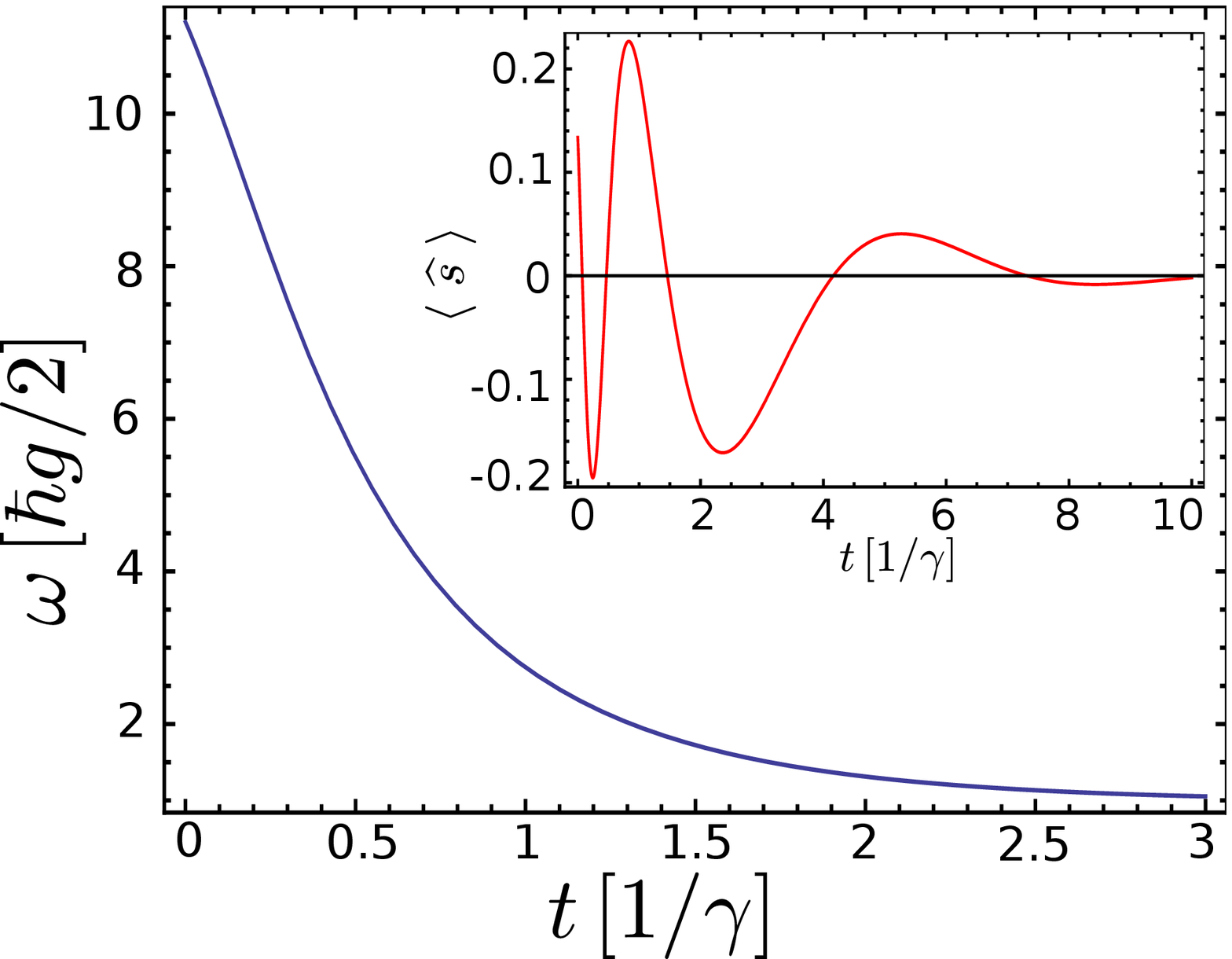,width=2.5in,clip=}}
 \caption[figomega2]{Time-dependence of the effective frequency $\omega(t)$ and of the magnetisation $m(t)=\left< \wht{s}\,\right>(t)$, 
 in the weak-quantum-coupling regime with $g=0.1 J/\hbar^2$, for different ratios $A/N$: \\
 \underline{Left panel}: $A=0.1$, $N=10$ 
 \underline{Centre panel}: $A=4$, $N=5$  
 \underline{Right panel}: $A=10$, $N=0.1$.  
 In all cases, $\omega(t)$ decays monotonously, in analogy with the case $A=N$, see left panel in fig.~\ref{fig:omega_ana}. 
 }
\label{fig:weak}
\end{figure}

The same two regimes also arise when $A\neq N$. In fig.~\ref{fig:weak}, the behaviour in the weak-quantum-coupling
regime is illustrated for choices such that either $A\ll N$, $A \simeq N$ or $A \gg N$, respectively. In this regime, we find qualitatively 
the same behaviour already shown in fig.~\ref{fig:omega_ana} for $A = N$: the effective frequency $\omega(t)$ 
decays monotonously (almost exponentially) and
the decay of the magnetisation is a simple damped oscillation, of which the frequency decreases, towards $\omega(\infty) = \hbar g /2$.

\begin{figure}
 \centerline{
 \psfig{figure=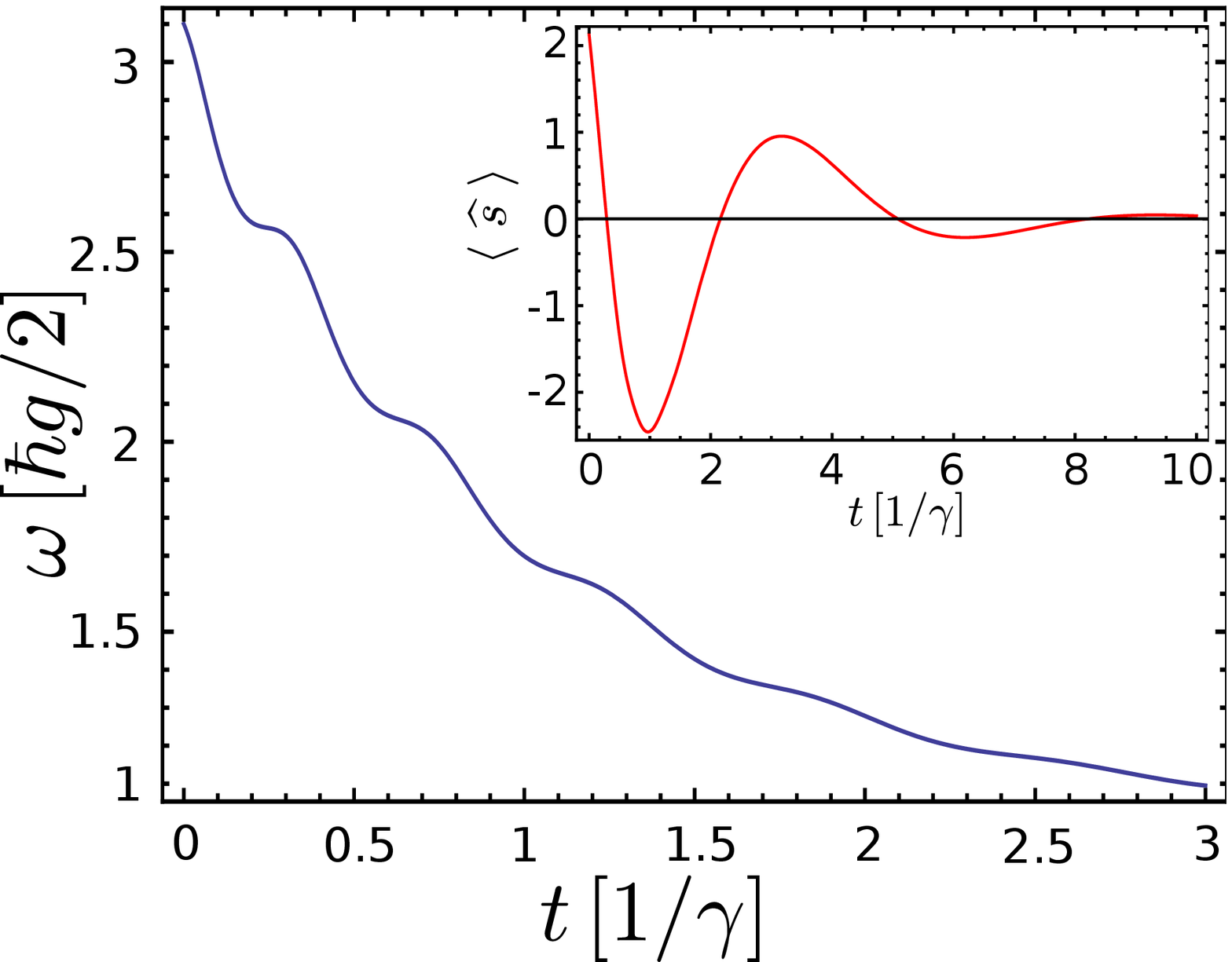,width=2.5in,clip=} 
 ~~\psfig{figure=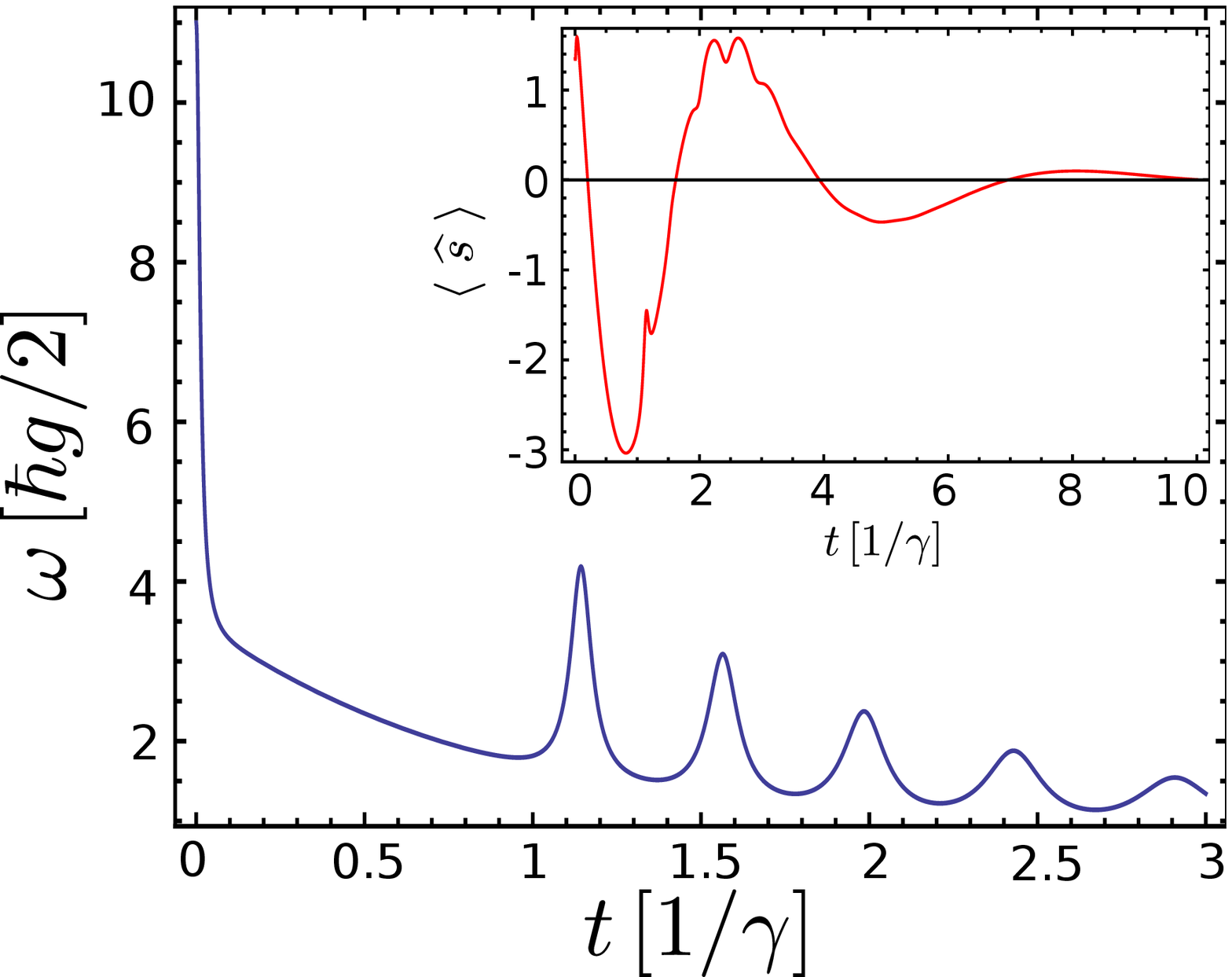,width=2.5in,clip=}
 \psfig{figure=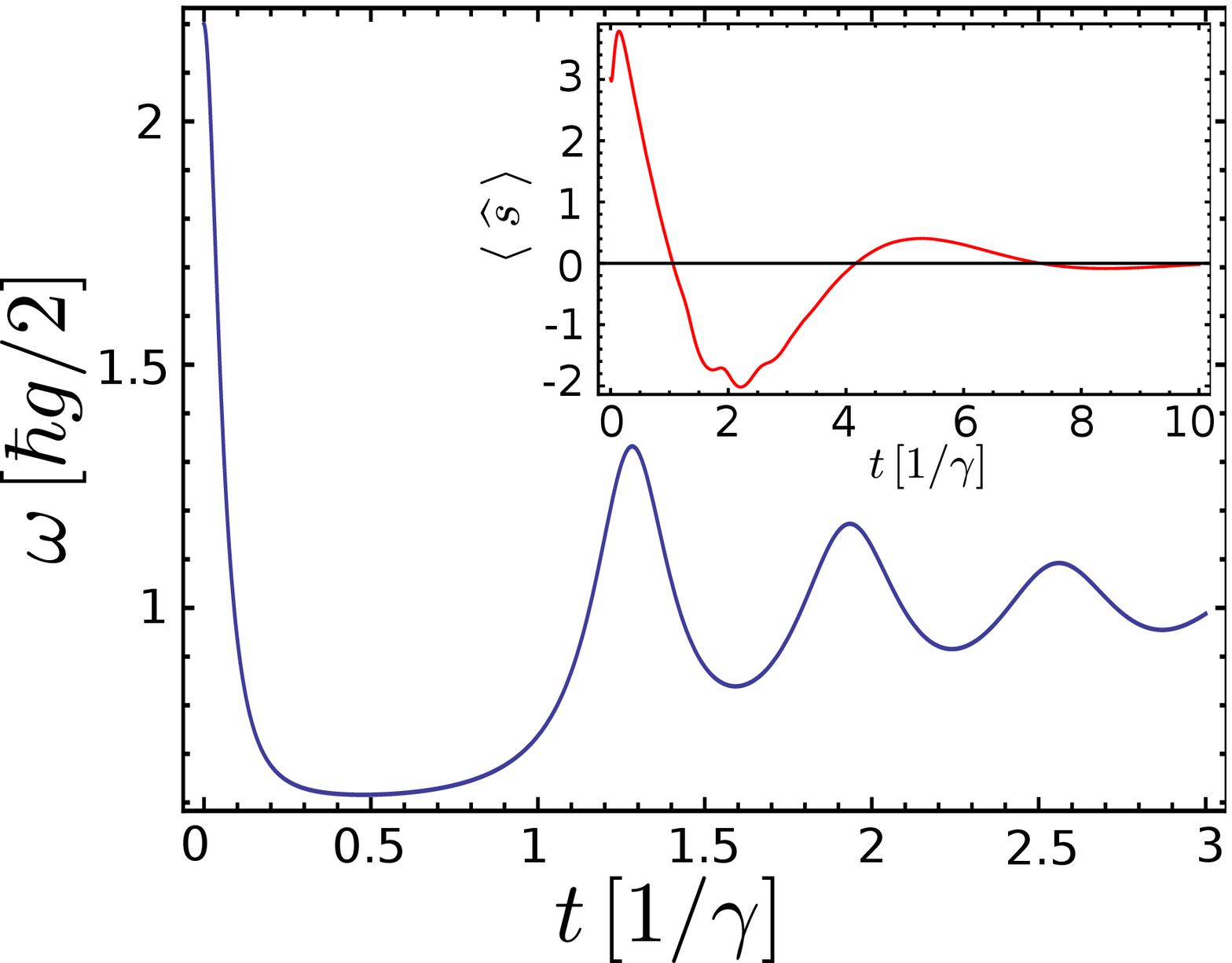,width=2.5in,clip=}}
 \caption[figomega3]{Time-dependence of the effective frequency $\omega(t)$ and of the magnetisation $m(t)=\left< \wht{s}\,\right>(t)$, 
 in the strong-quantum-coupling regime for the following parameters: \\
 \underline{Left panel}:  $A=0.1$, $N=10$ and $g=7 J/\hbar^2$, 
 \underline{Centre panel}: $A=4$, $N=3$, $g=10 J/\hbar^2$, 
 \underline{Right panel}:  $A=1$, $N=0.1$, $g=10 J/\hbar^2$. 
 }
\label{fig:strong}
\end{figure}

Fig.~\ref{fig:strong} displays the behaviour in the strong-quantum-coupling
regime, again for different choices such that  either $A\ll N$, $A \simeq N$ or $A \gg N$, respectively. 
When $A\ll N$, quantum effects, after a rapid initial drop,  merely lead to
a small modulation of an essentially still monotonic decay of $\omega(t)$, 
which in turn is not very visible in the oscillating decay of the 
magnetisation, see the left panel in fig.~\ref{fig:strong}. 
On the other hand, quantum effects do become much more pronounced whenever
$A\gtrsim N$. After a clearly visible drop in $\omega(t)$ at short times, 
followed by a monotonous decay up to times $t\sim {\rm O}(1/\gamma)$, 
strong peaks overlay the background evolution. These are also visible
in the relaxation behaviour of the magnetisation, 
where a secondary periodic behaviour appears, see centre and right panels in fig.~\ref{fig:strong}.
This is qualitatively analogous to the right panel in fig.~\ref{fig:omega_ana}.

In order to better appreciate the r\^ole of the spherical constraint, 
let us recall the well-known behaviour of a quantum harmonic
oscillator without it \cite{Breu02,Carm99,Scha14}, as was also encountered in section~2 for the 
single-mode Dicke model. The Hamiltonian is again taken to be given by (\ref{2.4}), 
with the fixed frequency $\omega=\omega_h=\mbox{\rm cste.}$. Upon coupling the system to a thermal bath, the dynamics is again
described by the Lindblad equation (\ref{Gl:Lindblad}). From this, the equation of motion for 
$\left<\down\right>$ is rapidly written down, being the
analogue of (\ref{eq:a}), and solved \cite{Breu02,Carm99,Scha14}. It follows that the average magnetisation has the form 
\BEQ
\left<\wht{s}\, \right>  = e^{-\demi \gamma t}\left( a \cos \omega_h t + b \sin \omega_h t \right)
\EEQ
where $a,b$ are constants. One has a regular damped oscillation, with fixed frequency 
$\omega=\omega_h$. The distinct regimes of weak and strong
quantum couplings seen in figure~\ref{fig:omega_ana} do not appear. 
Although the long-time limit looks to be analogous to the one derived in 
eq.~(\ref{gl:thetainf}) the finite-time behaviour of the single-spin spherical model 
allows for considerable more complexity. For example, even in
the weak-coupling regime, the decrease of the oscillation frequency $\omega(t)$ 
is clearly visible in the non-harmonic oscillations of the magnetisation
in the inset of the left panel in figures~\ref{fig:omega_ana}, \ref{fig:weak} and \ref{fig:strong}.

\section{Steady-state solution in the mean-field description}
\label{ssec:sss0field}

In this section, we consider the single {\sc sqs} at $T=0$ as a mean-field 
approximation of an $N$-body problem. In the most simple mean-field scheme of magnetic phase transitions, 
one replaces the spin-spin interactions by an effective external 
magnetic field $B=B_{\rm eff}$ \cite{Jaeg96}, which is then self-consistently related to the magnetisation, 
by writing $B_{\rm eff} = \kappa \left< \hs\, \right>$ 
with some appropriately chosen proportionality constant $\kappa$. 

We formally keep the above Lindblad equation (\ref{Gl:Lindblad}) for the description of the dynamics, even if $B\ne 0$. 
Our main interest will be the determination of the structure of the phase diagram which means that essentially, we are going
to look at the stability of the disordered phase with a vanishing magnetisation. In principle, the Lindblad operators $L_{\alpha}$ no longer couple
directly to the eigenmodes of the system, and one cannot expect a relaxation to an equilibrium state \cite{Breu02}. 
Rather, the relaxation should be towards some non-equilibrium steady-state ({\sc ness}) whose properties we are going to study. 
On the other hand, since we are mainly interested in the regime $B_{\rm eff} \sim \left< \hs\, \right> \ll 1$, these differences should 
not be very large. Also, in the quantum spherical model one expects that a mean-field approximation should correctly describe the quantum 
critical behaviour at $T=0$ above the upper critical dimension, $d>d^*=3$ \cite{Sach99,Taeu14}. 

\subsection{Zero-temperature phase diagram}

In order to start with the zero-temperature case ($T=0$), we introduce the definitions
\BEQ
x_1 := \Re\left< \down\, \right>, \ 
x_2  := \Im\left< \down\, \right>, \ 
x_3  := \Re\left<\down \down\, \right>, \ 
x_4  := \Im\left<\down \down\, \right>, \ 
x_5 := \left<\up \down\, \right>
\EEQ
and find from eqs. (\ref{eq:aa}), (\ref{eq:adaggera}) and (\ref{eq:a}) 
the following set of real-valued equations of motion of the {\sc sqs} in an external magnetic field $B$
\vspace{-.5cm}
\begin{center}
\begin{minipage}[h]{.49\textwidth}
\BEA
\label{eq:x1}
\dot{x}_1 &=& -\frac{\gamma}{2} x_1 + \omega x_2
\\
\label{eq:x2}
\dot{x}_2 &=& -\frac{\gamma}{2} x_2 - \omega x_1 +\demi \sqrt{\frac{2g}{\hbar\omega}}{B}
\\
\label{eq:x3}
\dot{x}_3 &=& -\gamma x_3 +2\omega x_4 -\sqrt{\frac{2g}{\hbar\omega}}B x_2 
\EEA
\end{minipage}
\begin{minipage}[h]{.49\textwidth}
 \BEA
 \label{eq:x4}
\dot{x}_4 &=&  -\gamma x_4 - 2\omega x_3 +\sqrt{\frac{2g}{\hbar\omega}}B x_1\\
\label{eq:x5}
\dot{x}_5 &=& -\gamma x_5 +\sqrt{\frac{2g}{\hbar}} x_2
 \EEA
\end{minipage}
\end{center}
We now cast this system of equations as a self-consistent mean-field approximation by relating the external field $B$ to the magnetisation, 
viz. $B=\kappa \left<\hs\,\right>$. 
Then, recall (\ref{2.3}) and also use the spherical constraint (\ref{eq:constraint}) 
in order to eliminate the variable $x_5$. We wish to analyse the stationary state, for which we have the system of equations
\vspace{-.5cm}
\begin{center}
\begin{minipage}{.48\textwidth}
\BEA
0&=& -\frac{\gamma}{2} x_1 + \omega x_2\\
0 &=& -\frac{\gamma}{2} x_2 - \omega x_1 +g\kappa\frac{x_1}{\omega}\\
0 &=& -\gamma x_3 +2\omega x_4 -2g\kappa\frac{x_1x_2}{\omega}
\EEA
\end{minipage}
\begin{minipage}{.51\textwidth}
\BEA
  0 &=&  -\gamma x_4 - 2\omega x_3 +2g\kappa\frac{x_1^2}{\omega}\\
  0 &=& -\gamma\frac{\omega}{\hbar g} + \gamma x_3 + \frac{\gamma}{2} + 2g\kappa \frac{x_1x_2}{\omega}~
\EEA
\end{minipage}
\end{center}
with the five independent variables $x_1$, $x_2$, $x_3$, $x_4$ and $\omega=\omega(\infty)$.

This system has two distinct solutions for $\omega$: one corresponds to a {\em disordered state}, labelled $\text{d}$, 
with frequency $\omega_{\rm d}=\demi \hbar g$ and $x_1=x_2=x_3=x_4=0$
and the other one corresponding to a magnetically {\em ordered state}, labelled $\text{o}$, with frequency $\omega_{\rm o}^2=g\kappa -\gamma^2/4$ and
the $x_1,\ldots,x_4$ non-vanishing. Compactly, the two physically distinct stationary states can be distinguished by their frequencies
\BEQ
\mbox{\rm disordered:~} \omega_{\text{d}} = \frac{\hbar g}{2} \;\;\; ; \;\;\; 
\mbox{\rm ordered:~} \omega_{\text{o}} = \sqrt{g\kappa -\frac{\gamma^2}{4}} \ .
\EEQ

\begin{figure}[tb]
\centerline{\psfig{figure=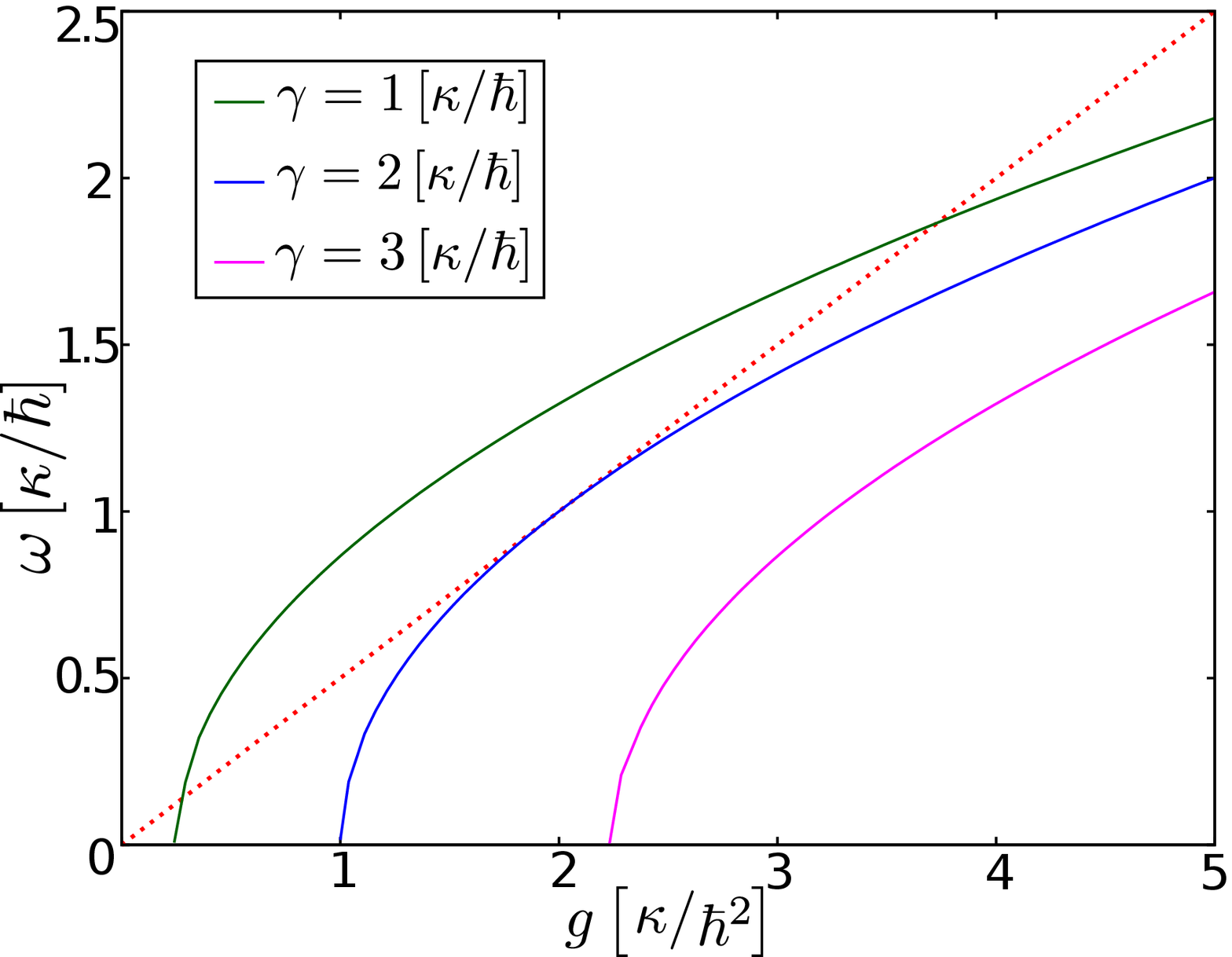,width=3.5in,clip=} ~~\psfig{figure=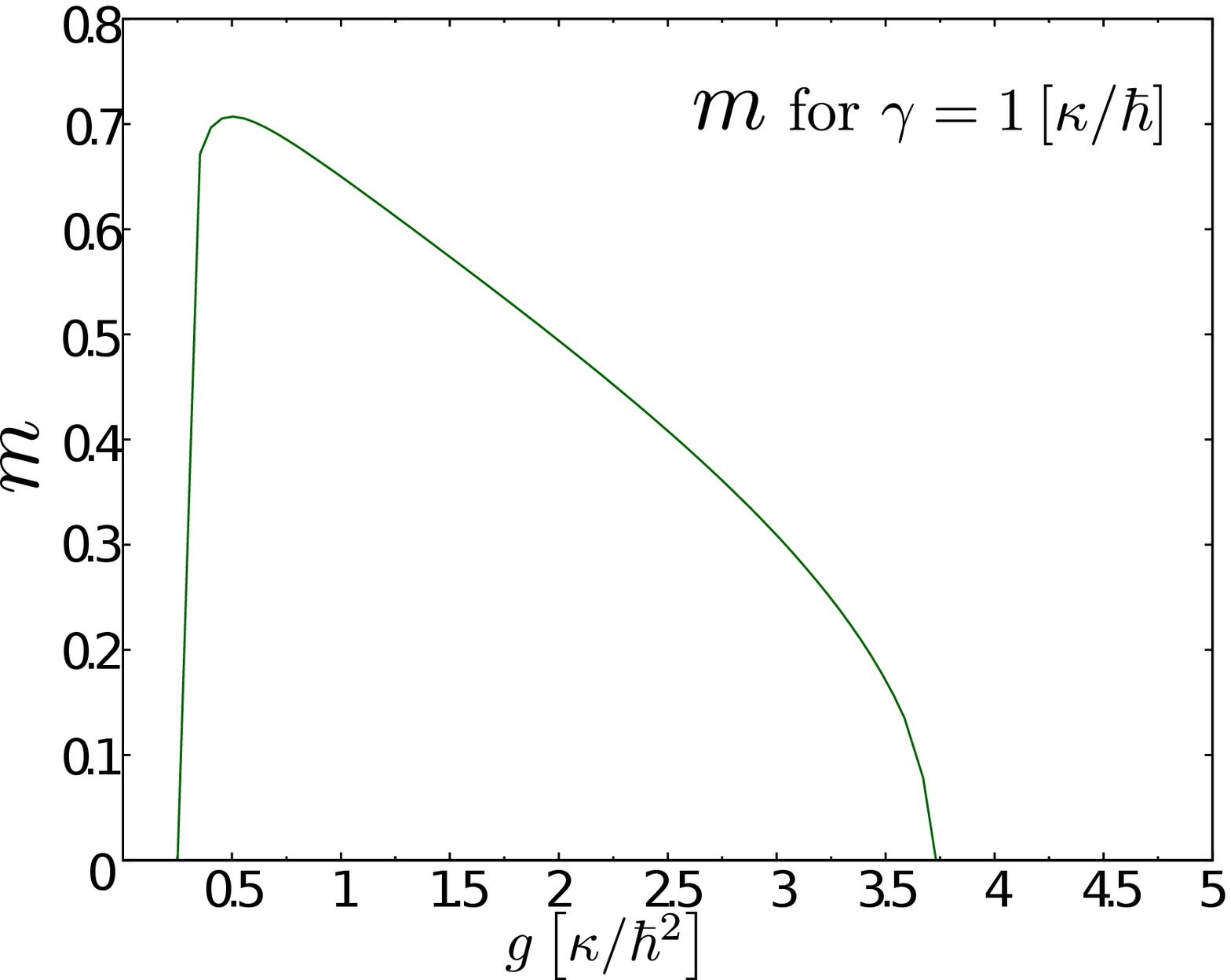,width=3.5in,clip=}}
\caption[fig2]{Left panel: stationary frequency $\omega$ as a function of the quantum coupling $g$ for the two distinct 
steady-state solutions, for several values of $\gamma$, and for temperature $T=0$.The full lines give $\omega_{o}$, the dotted
line $\omega_d$. \\
Right panel: average magnetisation $m$ as a function of $g$, for $\gamma=1$ and $T=0$. 
There are two quantum critical points for $\gamma < 2 \kappa/\hbar$  and one multi-critical point for 
$\gamma = 2\kappa/\hbar$.
\label{fig:omega}}
\end{figure}

In the left panel of fig.~\ref{fig:omega}, we characterise the different stationary states by displaying the stationary
frequencies $\omega$ as a function of the quantum coupling $g$, for several values of the dissipation coupling $\gamma$. 
The red (dotted) straight line corresponds to the disordered solution $\omega_{\text{d}}$, 
while the other lines correspond to the ordered solution $\omega_{\text{o}}$, for different values of the damping $\gamma$. 
Depending on the value of $\gamma$, we either find 
\begin{itemize}
  \item two intersections, for $ 0 < \gamma < 2 \kappa / \hbar$, at
  \BEQ\label{eq:critline}
  g_{1,2} = 2 \frac{\kappa}{\hbar^2} \mp \sqrt{4\frac{\kappa^2}{\hbar^4} - \frac{\gamma^2}{\hbar^2}}
  \EEQ
  \item one intersection, for $\gamma = 2\kappa/\hbar$, at 
  \BEQ
  g_{c} = 2 \frac{\kappa}{\hbar^2}
  \EEQ
  \item no intersection, for $2 \kappa/\hbar < \gamma$.
\end{itemize}
It turns out that the larger of these solutions $\omega$ is stable, in the sense 
of a linear stability analysis of the system (\ref{eq:x1}-\ref{eq:x5}), as shown in appendix~C. 
In other words, whenever no intersections between $\omega_{\rm o}$ and $\omega_{\rm d}$ occur, the disordered 
solution, with frequency $\omega_{\text{d}}$, is always stable and the ordered solution, with frequency $\omega_{\text{o}}$, is always unstable. 
On the other hand, in the case of two intersections, the
disordered solution, with frequency $\omega_{\text{d}}$, is only stable if either $g<g_1$ or $g>g_2$, while the ordered solution
$\omega_{\text{o}}$ is stable in the intermediate region $g_1<g<g_2$. In this intermediate regime, there is a non-vanishing
spontaneous magnetisation 
\BEQ
m^2 = \left< \hs\,\right>^2 = 
\frac{\gamma^2}{4\kappa g}\left( 1 -\frac{\hbar g}{2\omega} \right)\left( 1 + \frac{4\omega^2}{\gamma^2}\right) \ ,
\EEQ
whose dependence on $g$, for a fixed value $\gamma=1$, is shown in the right panel of figure~\ref{fig:omega}. 
This makes apparent the physical origin of the labels `ordered' and `disordered'. 
Two distinct quantum phase transitions occur at $g_1$ and 
$g_2$, respectively. Near these quantum critical points, we can rewrite the magnetisation as follows, with $j=1,2$
\BEQ
m^2 \approx \frac{\gamma}{4\kappa g_j^2}\left( 1+\frac{\hbar^2g_j^2}{\gamma^2}\right)
\frac{\sqrt{4\kappa^2-\gamma^2\hbar^2}}{2\kappa-\sqrt{4\kappa^2-\gamma^2\hbar^2}}\left|g-g_j \right| \ .
\EEQ
Recalling the standard definition of the magnetisation critical exponent, $m^2\sim |g-g_j|^{2\beta}$, we read off
the expected mean-field value $\beta=1/2$. 

\begin{figure}[tb] 
\centerline{\psfig{figure=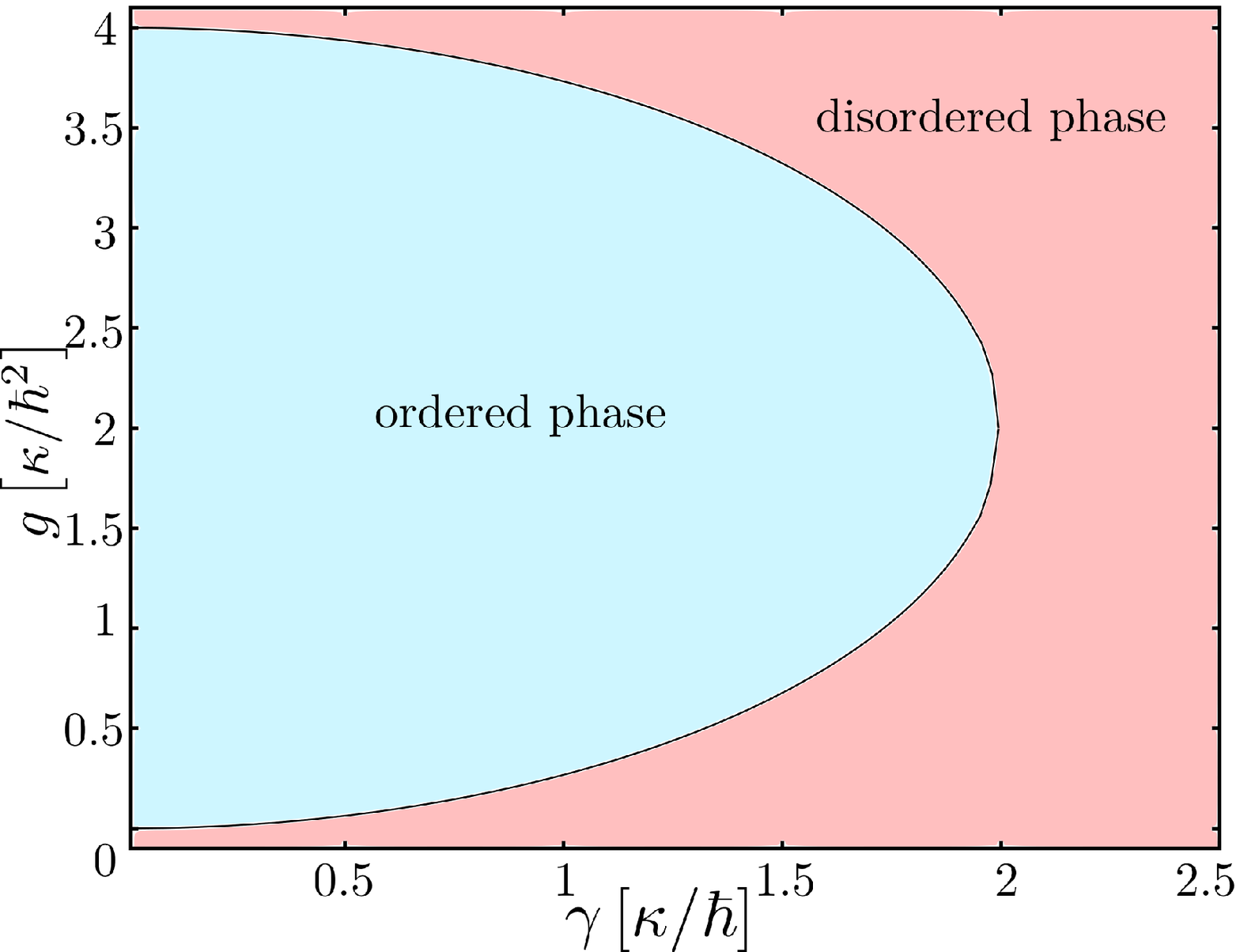,width=3.5in,clip=}~~\psfig{figure=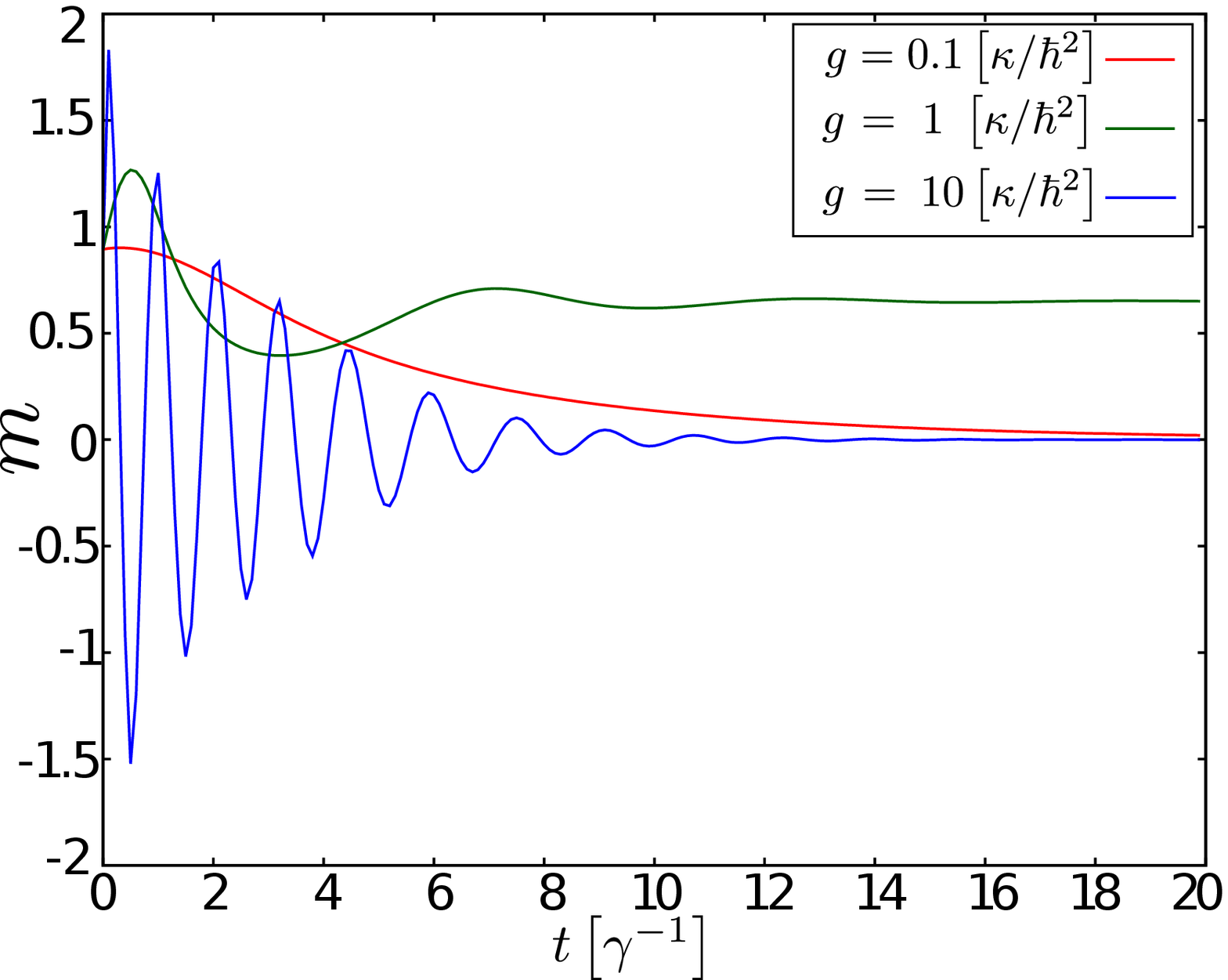,width=3.5in,clip=}}
\caption[fig2]{\underline{Left panel:} mean-field phase diagram of the {\sc sqs} at $T=0$
with its two distinct phases. The critical line is a parabola, given by eq.~(\ref{eq:critline}). 
\underline{Right panel:} relaxation of the magnetisation, along the line 
$\gamma =1$. In the disordered phase with $g<g_1$, there is a monotonous exponential decay (red curve), 
in the disordered phases with $g>g_2$, there is an oscillatory decay (blue curve). 
In the intermediate phase $g_1<g<g_2$, there is
a relaxation towards a magnetically ordered stationary state (green curve). 
}
\label{fig:phase}
\end{figure}

The mean-field phase diagram is shown in the left panel of fig.~\ref{fig:phase}. The ordered and the disordered phases
are clearly separated. For sufficiently large values of the damping coupling $\gamma$, 
any ordered structure is simply dissipated away, for all values of the quantum coupling $g$. 
Analogously, for sufficiently large values of $g$, quantum disorder destroys
any magnetic order. Surprisingly, we find a re-entrance of the disordered phase also when the quantum coupling becomes
small enough~! This means that in order to have an ordered stationary state, a cooperative effect between the quantum
fluctuations, parametrised by $g$ and the dissipation, parametrised by $\gamma$, is required. This is a highly
non-intuitive effect of the coherent quantum dynamics, without an analogue in the classical spherical model. 

The distinction between the two regions of the disordered phase is further illustrated 
through the relaxation of the magnetisation, see the right panel of figure~\ref{fig:phase}. 
Although the stationary magnetisation always vanishes in the disordered
phase, the approach to this stationary value depends on value of the quantum coupling $g$. If $g>g_2$ is large enough, there
are magnetisation oscillations while for $g<g_1$ small enough, the approach towards to stationary value is monotonous. 
Some magnetic oscillations are also seen for relaxations within the ordered phase.

To what extent could one take these results, interpreted as coming from a mean-field approximation, 
as a useful guide for more complex
systems with stronger fluctuation effects~? The pronounced difference in the shape 
of the magnetisation curve $m=m(g)$, near to
$g=g_1$ and $g=g_2$, respectively, might suggest that fluctuation effects might turn the continuous transition at $g=g_1$ into a first-order transition. 
Of course, it would be important to check if the presence of a disordered state for quantum couplings $0<g<g_1$ remains
valid beyond the mean-field approximation. 
However, since mean-field theory is considered as a reliable qualitative guide (and even quantitatively
for the critical behaviour of the spherical model in $d>3$ dimensions), it appears plausible that the qualitative 
features of the phase diagram figure~\ref{fig:phase} and the different types of relaxation behaviour 
could reflect more than an artefact of a simple approximation scheme. 
To answer this questions requires a solution of the Lindblad equation of a full $N$-body version of the
quantum spherical model in dimensions $d>1$ and we hope to return to this problem in the future.  

\subsection{Finite-temperature corrections}

\begin{figure}[t] \centering
\psfig{figure=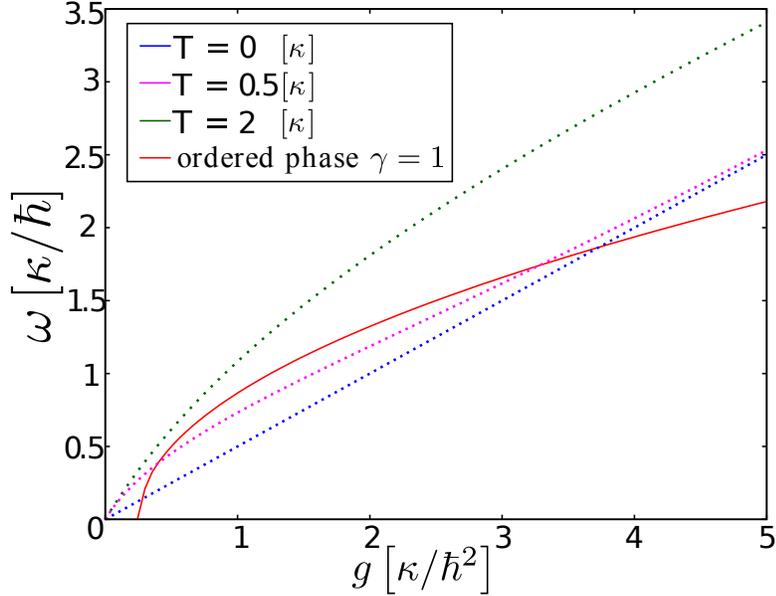,width=4in,clip=}
\caption[fig4]{
Effective frequencies $\omega$ of the different solutions of the steady-state, 
for a small temperature $T>0$ and $\gamma=1$, as a function of $g$. 
The frequency $\omega_{\rm o}$ of the ordered phase is given by the full red line. The frequencies $\omega_{\rm d}$ for the disordered
phase is given by the dotted blue, magenta and green lines for $T/\kappa=[0,0.5,2]$, from bottom to top. 
}
\label{fig:thermal}
\end{figure}

For a sufficiently small temperature $T> 0$, we shall calculate the first-order correction to the above zero-temperature solution by 
expanding the occupation number for $T\ll \hbar \cdot \min_{t\geq 0} \omega (t)$ 
\BEQ
n_{\omega} = \left(e^{\hbar \omega/T} - 1\right)^{-1} \simeq e^{-\hbar \omega/T} \ .
\EEQ
Since the temperature $T$ enters explicitly only into the the average $\left< \up \down \right>$ via the equation of motion (\ref{eq:adaggera}),
if follows that the the ordered zero-temperature solution and its frequency $\omega_{\rm o}$ remains unaffected by the temperature, to leading order, 
while the frequency of the disordered solution is slightly shifted, according to 
\BEQ \label{eq:thermal}
 \omega_{\text{d}}(T,g) \approx \frac{\hbar g }{2}+ \frac{T}{\hbar}W\left(\frac{\hbar^2 g}{T}
 e^{-\frac{\hbar^2g}{2T} }\right) \ .
\EEQ
where $W$ denotes the Lambert-$W$ function \cite{Abra65}. In figure~\ref{fig:thermal}, we compare the the frequency $\omega_{\rm o}$ 
of the ordered phase with the temperature-dependent frequencies $\omega_{\rm d}$ of the disordered state. As the temperature $T$ increases, the
curves of $\omega_{\rm d}(g)$ bend downwards but provided $T$ does not grow too large, one can still find two intersections. 
This indicates that the topology of the phase diagram (left panel of figure~\ref{fig:phase}) should remain unchanged for a sufficiently small temperature 
$T>0$ such that the two quantum phase transitions obtained at $T=0$ persist.

\section{Conclusions}

We have studied the coherent quantum dynamics, as described by a Lindblad equation, of a simple toy model, consisting of a single
quantum oscillator which was also assumed to obey a constraint analogous to the quantum spherical model of ferromagnetism. While the
low-energy modes of that model look very similar to the ones of the Dicke model (in the rotating-phase approximation), an essential
difference arises from the effective time-dependence of the frequency $\omega(t)$, as determined from the spherical constraint, while in the
Dicke model, the frequency is usually taken to be a constant. Our aim has been to understand better the phenomenological consequences 
of describing a coherent quantum dynamics of an open quantum system, coupled to a bosonic heat bath via the Lindblad equation. 
We have found: 
\begin{enumerate}
\item the exact time-dependent solution, without an external field and at zero temperature, 
allows to distinguish two distinct relaxational regimes, of weak and of strong
quantum-coupling, respectively. In the weak-quantum-coupling regime, the relaxation is dominated by the dissipation, as described by the
dissipation coupling $\gamma$, whereas in the strong-quantum-coupling regime, intrinsic quantum oscillations lead to a more complex
phenomenology of the relaxation of physical observables, such as the magnetisation, see figure~\ref{fig:strong}. 
\item when considering our single-spin model, in the presence of an external magnetic field, as an effective mean-field approximation of a 
quantum ferromagnet, the stationary state displays a surprising structure of its phase diagram. 
Remarkably, it turns out that a magnetically ordered
state can only arise if both quantum disorder, parametrised by the coupling $g$, as well as dissipation, parametrised through $\gamma$, are present. 
For fixed values of $\gamma$ (not too small and not too large) we find {\em two} distinct quantum phase transitions at couplings $g_{1,2}$, such
that an ordered magnetic state is stable for couplings $g_1 < g<g_2$ and is unstable otherwise, see figure~\ref{fig:phase}.  

These zero-temperature quantum phase transitions are stable under a small thermal perturbation. 
\end{enumerate}
For the interpretation of these quantum phase transitions, recall that the quantum coupling $g$ plays for quantum phase transitions at $T=0$ 
in $d$ dimensions  a r\^ole analogous to the temperature $T>0$ in classical phase transitions in $d+1$ dimensions 
\cite{Sach99,Henk84a,Vojt96,Oliv06,Wald15}. Therefore, fixing a value of 
$\gamma<2\kappa/\hbar$ and looking at the phase diagram in fig.~\ref{fig:phase}, 
when increasing the value of $g$, starting from a small value $g\ll g_1$, we see that {\em increasing the quantum fluctuations leads to a magnetic
ordering of the system}. Only if $g>g_2$ becomes rather large, this order will melt again. In classical systems, 
this phenomenon is well-known
and was first observed in non-equilibrium steady-states \cite{Katz84,Helb00,Zia02,Borc14,Alta15} (and refs. therein), although it was pointed
out that it is not intrinsically a non-equilibrium effect \cite{Radz03} and simple examples of it are known even when detailed balance is maintained \cite{Zia02}. In the wide sense of order induced by larger fluctuations,
names such as  `{\em freezing-by-heating}' \cite{Helb00} or `{\em getting more from pushing less}' \cite{Zia02} have been invented for
this phenomenon, although such re-entrant behaviour has been known long before in equilibrium systems \cite{Radz03}. 
In non-equilibrium steady-states, this is related to the occurrence of {\em negative} responses. 
Freezing-by-heating was also observed experimentally in 
super-cooled water on negatively charged surfaces of the pyroelectric material LiTaO$_3$ \cite{Ehre10}. 
A negative response of the system's energy with respect to the bath temperature was also reported in the Dicke model \cite{Ross12}. 

Apparently we have observed a true quantum analogue of this well-known phenomenon,  
which one might call `{\em quantum order by quantum fluctuations}',  
since our control parameter is the quantum coupling $g$ and not the bath 
temperature $T$.\footnote{By analogy with \cite{Ross12}, 
this suggests that experimental observations of this effect could also use purely quantum control parameters and are not restricted to
purely thermodynamical variables. However, in the {\sc sqs} the average energy $\left<\wht{H}\right>$ of the stationary state 
increases monotonically with $g$. At the critical points $g=g_{1,2}$ the derivative $\partial\left<\wht{H}\right>/\partial g$ taken from the left
is clearly smaller than its analogue taken from the right.}
A common feature of systems undergoing freezing-by-heating
or their quantum analogue is that their Hamiltonians conserve the total number of quasi-particles 
\cite{Katz84,Helb00,Zia02,Ross12,Borc14}.\footnote{Since there is no obvious breaking of a symmetry between a macroscopic number
of ground states, there is no immediate relationship to the well-known `order-by-disorder' phenomenon, see \cite{Vill80,Berg07} and
\cite{Ross14} for an experimental observation in the pyrochlore magnet Er$_2$Ti$_2$O$_7$.}

Can one extend this observation in quantum systems, going beyond a simple mean-field scheme, 
towards a larger number of degrees of freedom~? A comparison with non-perturbative methods in field theory, with the view of a possible
collective-state interpretation or else with other examples of coherent quantum dynamics might lead to new insights. 
The answer to this question is left for future work \cite{Wald16}. 

\zeile{1}
\noindent 
{\bf\large Acknowledgements:} 
It is a pleasure to thank G.T. Landi, R. Betzholz, D. Karevski, 
A. Faribault, T. Gourieux, J.D. Noh, R.K.P. Zia and G. Morigi for useful discussions. 
This work was  partly supported by the Coll\`ege Doctoral franco-allemand Nancy-Leipzig-Coventry
({\it `Syst\`emes complexes \`a l'\'equilibre et hors \'equilibre'}) of UFA-DFH. 
SW is grateful to UFA-DFH for financial support through grant CT-42-14-II.

\newpage

\appsection{A}{Phenomenological dynamics in the quantum spherical model}

The so-called `spin-anisotropic quantum spherical model' ({\sc saqsm}) is defined, on a $d$-dimensional hyper-cubic lattice, 
and for a vanishing magnetic field $B=0$, by the Hamiltonian \cite{Wald15}
\BEQ \label{A1}
\wht{H} = \sum_{\vec{n}}\left(\frac{g}{2} \hp_{\vec{n}}^{\,2}+
\mathscr{S}J\,\hs_{\vec{n}}^{\,2} - \sum_{j=1}^d \left(
\frac{1+\lambda}{2}J\, \hs_{\vec{n}}\hs_{\vec{n}+\vec{e}_j}+
\frac{1-\lambda}{2} \frac{g}{2\mathscr{S}} \hp_{\vec{n}}\hp_{\vec{n}+\vec{e}_j}\right)
\right)
\EEQ
with the commutation relations $\left[\hs_{\vec{n}}, \hp_{\vec{m}}\right] = \II \hbar \delta_{\vec{nm}}$. 
The spherical parameter $\mathscr{S} : = \mu/2J$ is found self-consistently from the (mean) spherical constraint
$\left< \sum_{\vec{n}} \hs_{\vec{n}}^{\,2} \right> = {\cal N}$, where ${\cal N}$ is the number of lattice sites. The model's
parameters are $g,\lambda$ and $J$ (we shall almost always re-scale to $J=1$). 
Because of the symmetry in $\lambda$ \cite{Wald15}, we restrict throughout to $\lambda>0$. 
The usual quantum spherical model \cite{Ober72} is the special case $\lambda=1$. 
At equilibrium, for all $\lambda\ne 0$ and $d>1$, the model
has a continuous quantum phase transition at temperature $T=0$. 
The associated exponents and universal amplitude ratios are $\lambda$-independent, as expected from universality \cite{Wald15}. 
Remarkably, for dimensions $1<d\lesssim 2.065$, that phase transition is re-entrant in the sense that the critical
coupling $g_c=g_c(\lambda)$ is a non-monotonous function of $\lambda$ \cite{Wald15}. The {\sc saqsm} therefore allows to
analyse non-trivial quantum effects on its critical behaviour. Here, we shall show that {\it if the dynamics is taken
to be the analogue of the phenomenological `quantum Kramers equation' (\ref{1.3}), the system's behaviour becomes equivalent to
the classical case $g=0$, $\lambda=1$ for sufficiently large times.} 

\noindent {\bf Step 1:} Generalising the procedure leading to (\ref{1.3}) 
to generic values of $\lambda$, we find the `quantum Kramers equations' of motion (with $J=1$) \cite{Tail06,Dura15}
\BEA
\partial_t \hs_{\vec{n}} &=& g \hp_{\vec{n}} - \frac{1-\lambda}{2}\frac{g}{2\mathscr{S}(t)} 
\sum_{j=1}^d \left( \hp_{\vec{n}-\vec{e}_j} + \hp_{\vec{n}+\vec{e}_j} \right)
\label{A2} \\
\partial_t \hp_{\vec{n}} &=& -2 \hs_{\vec{n}} + 
\sum_{j=1}^d \left(\frac{1+\lambda}{2}  \left( \hs_{\vec{n}-\vec{e}_j}+\hs_{\vec{n}-\vec{e}_j}\right) +
\frac{1-\lambda}{2}\frac{g}{2\mathscr{S}(t)} \left( \hp_{\vec{n}-\vec{e}_j}+\hp_{\vec{n}-\vec{e}_j}\right)\right)
\nonumber \\
& &  -\gamma g \hp_{\vec{n}} + \wht{\eta}_{\vec{n}}
\label{A3}
\EEA
{\bf Step 2:} Using the Fourier representation (on a hyper-cubic lattice with ${\cal N}=N^d$ sites)
\BEA
\tilde{\hs}_{\vec{k}} = \sum_{\vec{n}}e^{\II\frac{2\pi}{N}\vec{k}\cdot\vec{n}}\, \hs_{\vec{n}} \;\; , \;\; 
\tilde{\hp}_{\vec{k}} = \sum_{\vec{n}}e^{\II\frac{2\pi}{N}\vec{k}\cdot\vec{n}}\, \hp_{\vec{n}} \;\; , \;\; 
\tilde{\wht{\eta}}_{\vec{k}} = \sum_{\vec{n}}e^{\II\frac{2\pi}{N}\vec{k}\cdot\vec{n}}\, \wht{\eta}_{\vec{n}}
\EEA
decouples the modes and brings the equations of motion to the form 
\BEQ \label{A5}
\partial_t \tilde{\hs}_{\vec{k}} = \frac{g}{\mathscr{S}(t)}\Lambda_2^2(t,\vec{k})\tilde{\hp}_{\vec{k}}
\;\; , \;\; 
\partial_t \tilde{\hp}_{\vec{k}} = -2\Lambda_1^2(t,\vec{k})\tilde{\hs}_{\vec{k}}-\frac{g\gamma}{\mathscr{S}(t)}
\Lambda_2^2(t,\vec{k})\tilde{\hp}_{\vec{k}} + \tilde{\wht{\eta}}_{\vec{k}}(t)
\EEQ
with the following eigenvalues of the Hamiltonian \cite{Wald15}
\BEQ\Lambda(t,\vec{k}) = \Lambda_1(t,\vec{k})\cdot \Lambda_2(t,\vec{k}) :=
\sqrt{\mathscr{S}(t) - \frac{1+\lambda}{2}\sum_{j=1}^d \cos k_j}\sqrt{\mathscr{S}(t) - 
\frac{1-\lambda}{2}\sum_{j=1}^d \cos k_j}
\EEQ
If one defines $\Omega(t,\vec{k}) := \exp \left(- \int_0^t \D\tau \frac{g\gamma }{\mathscr{S}(\tau)}\
\Lambda_2^2(\tau,\vec{k})\right)$ and denotes the convolution by $*$ (with respect to $\vec{k}$), 
the formal solution of (\ref{A5}) for the momenta reads 
\BEA
\tilde{\hp}_{\vec{k}}(t) = \tilde{\hp}_{\vec{k}}(0) \Omega(t,\vec{k})
+ \left(-2\Lambda_1(t,\vec{k})\tilde{\hs}_{\vec{k}}(t)+\tilde{\wht{\eta}}_{\vec{k}}(t)\right) * \Omega (t,\vec{k})
\EEA
(with a slight ab-use of notation concerning the convolution with respect to $\vec{k}$). 
Inserting this solution into the other eq. (\ref{A5}) for $\tilde{\hs}_{\vec{k}}$, we find in the long-time limit
\BEQ \label{A8}
\partial_t\tilde{\hs}_{\vec{k}}(t) \approx -\frac{1}{\gamma} \left( 
-\Lambda_1^2(t,\vec{k})\tilde{\hs}_{\vec{k}}(t)+\tilde{\wht{\eta}}_{\vec{k}}(t)\right) * \partial_t \Omega (t,\vec{k})\ ,
\EEQ
where we  dropped the contribution of the initial values of the conjugate momenta. 
For sufficiently large times, this is justified, since for $\lambda>0$, 
the spherical parameter $\mathscr{S}(t)\geq \frac{1+\lambda}{2}d$ \cite{Wald15}, 
hence $\Lambda_2^2(t,\vec{k})-\Lambda_1^2(t,\vec{k})\geq \lambda \sum_{j=1}^d \cos k_j\approx \lambda d$ 
in the low-momentum limit which is
relevant for the slowest modes. Therefore, the conjugate momenta decay at least as fast as 
\BD
\tilde{\wht{p}}(t) \sim \exp\left(-\frac{2\lambda}{1+\lambda} g\gamma\,t\right)=
\exp\left(-\frac{2\lambda}{1+\lambda}\frac{t}{t_{\rm deco}}\right)
\ED
whereas the decay of the slowest spin modes in the system is according to \typeout{*** saut de ligne ici ***} \\ 
$\tilde{\hs}(t)\sim \exp\left( - \Lambda_1^2(t,\vec{k})/\gamma\right)
\sim \exp\left(- \left[\mathscr{S}(t)-(1+\lambda)d/2\right]/\gamma +{\rm O}(\vec{k})^2\right)$. 

\noindent{\bf Step 3:} We want to show how, in the long-time limit, the equation of motion (\ref{A8})  
reduces to the Langevin equation of the classical spherical model, with $g=0$ and $\lambda=1$. 

In order to extract the leading long-term behaviour (\ref{A8}), 
one first maps this differential equation to an algebraic one, by using the Laplace transform
\BEQ
^{\star}\tilde{\hs}_{\vec{k}}(u):= \mathcal{L}\left(\tilde{\hs}_{\vec{k}}(t)\right)(u) 
= \int_0^{\infty} \!\D t\ e^{-u t}\, \tilde{\hs}_{\vec{k}}(t)
\EEQ
We find
\BEQ \label{eq:FL}
u \cdot ^{\star}\tilde{\hs}_{\vec{k}} (u) - \tilde{\hs}_{\vec{k}} (0) 
= \frac{1}{\gamma} \left(1 - u \cdot\ \hspace{-2mm} ^{\star}\Omega(u,\vec{k}) \right)
\cdot \mathcal{L}\left(-2\Lambda_1^2(t,\vec{k}) \tilde{\hs}_{\vec{k}}(t)+\tilde{\wht{\eta}}_{\vec{k}}(t)\right)(u) 
\EEQ
We shall need the large-time asymptotics of $\Omega(t,\vec{k})$, and its Laplace transform $^{\star}\Omega(u,\vec{k})$. 
Indeed, in the long-time limit, the integral in the exponential 
can be approximated by a time-independent average, as follows 
\BEA \nonumber
^{\star}\Omega(u,\vec{k}) &=& \int_0^{\infty} \!\D t\ e^{-ut} e^{-{\gamma} \int_0^t \D\tau \frac{g}{\mathscr{S}(\tau)} \Lambda_2^2(\tau,\vec{k})}
\:=\: \int_0^{\infty} \!\D t\ e^{-\left(u + t^{-1}\int_0^t\D\tau \frac{g\gamma}{\mathscr{S}(\tau)} \Lambda_2^2(\tau,\vec{k})\right) t}
\\
&\simeq& \int_0^{\infty} \!\D t\ e^{-\left(u + g\gamma\left<\frac{1}{\mathscr{S}} \Lambda_2^2(\vec{k})\right>\right) t}
\:=\: \left( u + g\gamma\left< \frac{1}{\mathscr{S}}\Lambda_2^2(\vec{k}) \right> \right)^{-1}
\EEA
where we define the average 
$\left<\frac{1}{\mathscr{S}} \Lambda_2^2(\vec{k})\right> := t^{-1} \int_0^t \!\D\tau\, \frac{1}{\mathscr{S}(\tau)} \Lambda_2^2(\tau,\vec{k})$. 
Inserting this result into eq. (\ref{eq:FL}), we find
\BEQ \label{A12}
u \cdot \ \hspace{-1mm}^{\star}\tilde{\hs_{\vec{k}}} (u)-\tilde{\hs}_{\vec{k}} (0) = \frac{1}{\gamma}  \frac{\left<\Lambda_2^2(\vec{k})/\mathscr{S} \right>}{\frac{1}{\gamma g} u +\left<\Lambda_2^2(\vec{k})/\mathscr{S} \right>}
\mathcal{L}\left(-2\Lambda_1^2(t,\vec{k})\tilde{\hs}_{\vec{k}}(t)+\tilde{\wht{\eta}}_{\vec{k}}(t)\right)(u) \ .
\EEQ
Standard Tauberian theorems \cite[ch. XIII.5]{Fell71} relate the asymptotic long-time behaviour of a function $f(t)$ as $t\to\infty$ to the
behaviour of its Laplace transform ${\cal L}(f)(u)$ as $u\to 0^+$. Therefore, in order to find the leading 
long-time behaviour of the spin operators $\tilde{\wht{s}}_{\vec{k}}(t)$, we first analyse the leading $u\to 0^+$-behaviour of the 
expression (\ref{A12}), which gives
\BEA
u\cdot \ \hspace{-1mm}^{\star}\tilde{\hs}_{\vec{k}} (u)-\tilde{\hs}_{\vec{k}} (0) \simeq 
\frac{1}{\gamma} \mathcal{L}\left(-2\Lambda_1^2(t,\vec{k})\hs_{\vec{k}}(t)+\tilde{\wht{\eta}}_{\vec{k}}(t)\right)(u)
\EEA
and then, via an inverse Laplace transform, we find the sought effective long-time form of the equations of motion for the spin variables 
(herein, we also restore the coupling $J$)
\BEA\nonumber
\partial_t \tilde{\hs}(t,\vec{k}) &\simeq& - \frac{2J}{\gamma} \Lambda_1^2 (t,\vec{k}) \tilde{\hs}(t,\vec{k}) 
+ \frac{1}{\gamma}\tilde{\wht{\eta}}_{\vec{k}}(t)\\
&=&-\frac{2J}{\gamma}\left(\mathscr{S}(t) - \frac{1+\lambda}{2}\sum_{j=1}^d \cos k_j\right)\tilde{\hs}(t,\vec{k}) 
+\frac{1}{\gamma} \tilde{\wht{\eta}}_{\vec{k}}(t) 
\label{eq:Langevin} \ ,
\EEA
which we now compare to the Langevin equation of the classical spherical model \cite{Ronc78,Godr00b}. 

Indeed, if we take $\lambda=1$, we see that (\ref{eq:Langevin}) reproduces the classical Langevin equation if we choose 
\BEQ
{\gamma} = 2J \ .
\EEQ
and renormalise the temperature $T\mapsto \gamma T$ (unimportant for a quantum phase transition at $T=0$). 
Moreover, the result (\ref{eq:Langevin}) shows that the parameter $\lambda\ne 1$ merely gives rise to a renormalisation of the time $t$ and 
of the spherical parameter. Therefore, the supposed `quantum dynamics' (\ref{A2},\ref{A3}) does {\em not} relax to the equilibrium state of
the {\sc saqsm}-model (\ref{A1}), but rather dissipates away the non-trivial quantum effects 
\cite{Wald15} on the phase boundary $g_c=g_c(\lambda)$. 

\appsection{B}{Solution of equation (\ref{eq:phase})}
\label{app:annihilator}
In the main text, the phase $\Theta=\Theta(t)$ was shown to obey the equation (\ref{eq:phase}), which reads 
\BEQ \label{B1}
-\frac{\dot{\Theta}}{\hbar g} =2A\, e^{-\gamma t}\cos\Theta + 2N\, e^{-\gamma t} +1 \ .
\EEQ
and where $A$, $N$, $\gamma$ and $\hbar g$ are real constants. 
We shall solve this equation explicitly by mapping it to a well-known Riccati equation.

Let $y(t) := e^{\II\Theta(t)}$ and re-write eq.~(\ref{B1}) as
\BEQ
\frac{\II}{\hbar g} \dot{y} e^{\gamma t} = ye^{\gamma t} +2Ny + A\left(y^2+1\right)
\EEQ
A change of the time-scale, according to $\tau = e^{-\gamma t}$, together with the definitions
\BEQ \label{B3}
Y(\tau) := y(t), \hspace{.5cm} \mathcal{A} := \frac{\hbar g}{\II\gamma} A, \hspace{.5cm} 
B := \frac{2\hbar g N}{\II\gamma}, \hspace{.5cm} C := \frac{\hbar g }{\II\gamma} \ .
\EEQ
reduces this to the following {\em Riccati equation} 
\BEQ
\tau \dot{Y}(\tau) + \left(B\tau +C\right)Y(\tau) + \mathcal{A}\tau Y^2(\tau) + \mathcal{A}\tau = 0
\EEQ
which depends on the three parameters ${\cal A},B,C$. 
A standard method for solving this kind of equation consists in mapping it
to a second-order linear differential equation, by changing variables according to 
$\lambda Y =: \dot{u}/u$ \cite[sec. 1.2.2, eq. (45)]{Poly}. Hence  
\BEQ \label{B5}
\tau \ddot{u}+(B\tau + C)\dot{u}+\mathcal{A}^2 \tau u = 0 \ .
\EEQ
Next, simplify (\ref{B5}) by separating off an exponential, according to $u(\tau) = e^{-\kappa \tau}w(\tau)$, where $\kappa$ remains
to be chosen. We arrive at the following equation, for the unknown function $w(\tau)$
\BEQ \label{B6}
\tau \ddot{w} +\left[ C -(2\kappa-B) \tau\right]\dot{w} + \left[\tau (\kappa^2 - \kappa B + \mathcal{A}^2)-\kappa C\right] w = 0 \ .
\EEQ
We now choose the free parameter $\kappa$ in order to render the pre-factor of $w$ in eq.~(\ref{B6}) time-independent. This will
allow us to recognise (\ref{B6}) as a Kummer or Bessel differential equation. In principle, one might take either of the two possibilities
$\kappa = B/2 \pm \sqrt{B^2/4-\mathcal{A}^2}$. Without loss of generality, we prefer the choice $\kappa = B/2 + \sqrt{B^2/4-\mathcal{A}^2}$.
Eq.~(\ref{B6}) turns into the form 
\BEQ \label{B7}
\tau \ddot{w} +\left( C -\sqrt{B^2 - 4\mathcal{A}^2\,}\: \tau\right)\dot{w} +  \frac{C}{2}\left(B+\sqrt{B^2-4\mathcal{A}^2\,}\, \right) w = 0 
\EEQ
for which we have to distinguish two different cases. \\

\noindent 
\underline{{\bf 1.} Case $B/2 \neq \mathcal{A}$.} This case may stated alternatively by requiring $A\ne N$. We can define a rescaled
time variable $T = \tau \sqrt{B^2-4\mathcal{A}^2}$, which reduces (\ref{B7}) to a \emph{Kummer equation}
\BEQ \label{B8}
T \ddot{w} + (C-T)\dot{w} - \frac{C}{2}\left(1+\frac{B}{\sqrt{B^2-4\mathcal{A}^2}} \right)w = 0
\EEQ
A basis of solutions is given by Kummer's functions $M$ and $U$ \cite{Abra65}. The general solution of (\ref{B8}) is consequently
a linear combination of both
\BEQ
w(T) = K_1 U( \mathscr{T} ) + K_2 M (\mathscr{T}) \ ,
\EEQ
with the triplet of indices and the argument 
\BEQ
\mathscr{T} : = \left[-\frac{\hbar g}{2\gamma}\left( \II+\frac{1}{\sqrt{A^2/N^2 -1}}\right) \hspace{2mm} ; 
\hspace{2mm}
-\II\frac{\hbar g}{\gamma} \hspace{2mm} ; \hspace{2mm}
2\frac{\hbar g }{\gamma} \sqrt{A^2-N^2} e^{-\gamma t} \right]
\ .
\EEQ
Back-transformation to the required solution of the original first-order differential equation will introduce a relation between the 
two integration constants $K_1$ and $K_2$ of the second-order equation (\ref{B8}). 

Finally, we transform this result back to our original variable $\Theta=\Theta(t)$.
For this purpose, we introduce the shorthand
$ \mathscr{T}_{(x,y)} = \mathscr{T} + (x;y;0)$. The phase $\Theta$ then reads
\BEA \label{B11}
\cos\Theta(t) = \Re\hspace{-.2cm}\left(-\frac{N}{A}-\II\sqrt{1-\frac{N^2}{A^2}}+\frac{\hbar g}{\gamma} 
\left(\sqrt{1-\frac{N^2}{A^2}}-\II\frac{N}{A}\right)
\frac{\II\frac{\gamma}{\hbar g} K M(\mathscr{T}_{(1,1)})-U(\mathscr{T}_{(1,1)})}{K M(\mathscr{T})-U(\mathscr{T})}\right) ~~~
\EEA
Herein, the constant $K$ is related to the initial condition. 

\begin{figure}[ht]\centering
 \centerline{\psfig{figure=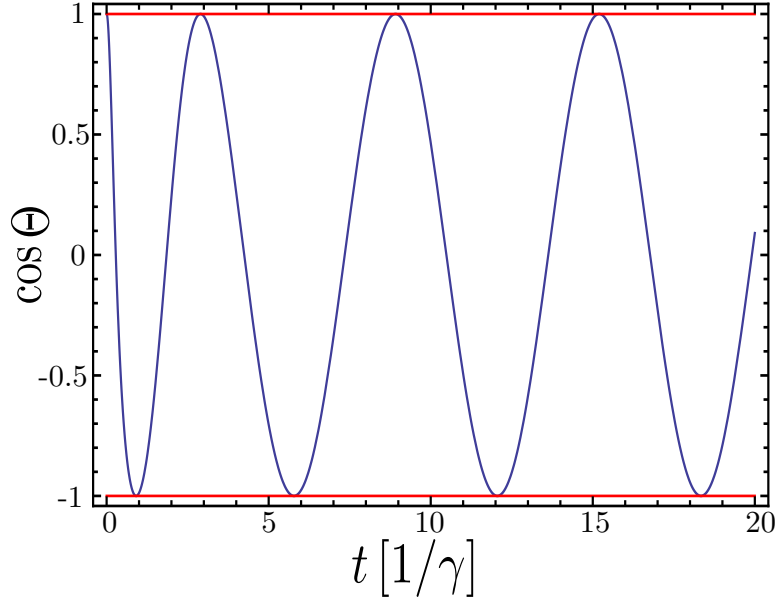,width=4in,clip=}}
\caption{Illustration of the right-hand-side of eq.~(\ref{eq:phase}), over against the time $t$, 
for the parameters $\hbar g/\gamma = 1$, $A=1$, $N=2$.}
\label{fig:sanity}
\end{figure}

As a `sanity check' of the whole procedure, we illustrate in figure~\ref{fig:sanity} an example of the right-hand-side of 
eq.~(\ref{B11}). It is satisfying to see that $\cos \Theta(t)$ assumes as values the full range $[-1,1]$, but does {\em not} exceed it, 
as it should be for a well-defined cosine function. This also holds true for all other parameter values. 

\noindent 
\underline{{\bf 2.} Case $B/2= \mathcal{A}$.} This case can also be specified by the condition $A = N$. 
Now, eq.~(\ref{B7}) turns into a Bessel differential equation 
\BEQ
\tau \ddot{w} +C \dot{w} +  \frac{BC}{2} w = 0 
\EEQ
with the general solution \cite{Boas}
\BEQ
w(\tau) = K_1 \tau^{(1-C)/2} J_{1-C}\left(\II\sqrt{2BC\tau} \right)+K_2 \tau^{(1-C)/2} J_{C-1}\left(\II\sqrt{2BC\tau} \right)
\EEQ
where $J_p(x)$ denotes the Bessel function of the first kind of order $p$ \cite{Abra65} ($C$ is not an integer, see (\ref{B3})). 
Transforming back to the original variables, we find 
\BEQ \label{B14}
\cos\Theta(t) = -\Re\left( 1 + \frac{\II}{\sqrt{A}}\, \e^{-\frac{\gamma}{2} t}
\frac{K J_{\II\frac{\hbar g}{\gamma}}\left(2\II\frac{\hbar g}{\gamma}\sqrt{A\, e^{-\gamma t}}\right)
-J_{-\II\frac{\hbar g}{\gamma}}\left(2\II\frac{\hbar g}{\gamma}\sqrt{A\, e^{-\gamma t}}\right)}
{K J_{1+\II\frac{\hbar g}{\gamma}}\left(2\II\frac{\hbar g}{\gamma}\sqrt{A\, e^{-\gamma t}}\right)
+J_{-1-\II\frac{\hbar g}{\gamma}}\left(2\II\frac{\hbar g}{\gamma}\sqrt{A\, e^{-\gamma t}}\right)}\right)
\EEQ
The constant $K$ is related to the initial condition. 
We also checked that the function $\cos\Theta(t)$ in (\ref{B14}) has the full range $[-1,1]$, 
in analogy with figure~\ref{fig:sanity}, as it should be. 

Eqs.~(\ref{B11},\ref{B14}) are the main result of this appendix and are quoted in the main text. 

\appsection{C}{Linear stability of the steady state}

We analyse the stability of the stationary solutions found in section \ref{ssec:sss0field} with a linear stability analysis.
Consider the equations of motion (\ref{eq:x1},\ldots,\ref{eq:x5}) and use the spherical constraint to eliminate the variable $x_5$. The
jacobian matrix $\cal J$ of the resulting system of five equations, in the variables $x_1,\ldots,x_4,\omega$, is 
\BEQ
\mathcal{J}= 
\begin{bmatrix}\vspace{.1cm}
-\gamma/2&\omega&0&0& \hbar g x_2\\\vspace{.1cm}
\kappa g/\omega - \omega&-\gamma/2&0&0&-\hbar g x_1\left( 1 + g\kappa/\omega^2 \right)\\ \vspace{.1cm}
-2g\kappa x_2/\omega&-2\kappa g x_1 /\omega&-\gamma&2\omega& 2 \hbar g\left( x_4 + \kappa g x_1x_2/\omega^2 \right)\\ \vspace{.1cm}
4\kappa g x_1/\omega & 0& -2\omega& -\gamma & -2\hbar g \left( x_3+\kappa g x_1^2/\omega^2 \right)\\ \vspace{.1cm}
0&0&0&2\omega& \hbar g\left( 2x_4-\gamma/\hbar g \right)
\end{bmatrix}
\EEQ
Inserting the disordered solution $\omega=\omega_{\rm d}$ gives the following list of eigenvalues $e_i$ of $\cal J$: 
\begin{center}
\begin{minipage}{.49\textwidth}
\BEA
e_1&=& -\gamma\\
e_2&=& -\gamma-\II\hbar g\\
e_3&=&-\gamma+ \II\hbar g
\EEA
\end{minipage}
\begin{minipage}{.49\textwidth}
\BEA
e_4&=&-\gamma/2-\sqrt{\kappa g -\hbar^2g^2/4}\\
e_5 &=&-\gamma/2+\sqrt{\kappa g -\hbar^2g^2/4}
\EEA
\end{minipage}
\end{center}
In the range $g \in (0,g_1)\cup (g_2,\infty)$, with $g_{1,2}$ given by (\ref{eq:critline}), 
all real parts of the eigenvalues are negative and thus the disordered solution is stable 
under small perturbations. On the other hand, for $g\in(g_1,g_2)$, the disordered solution is unstable, 
since the real part of the eigenvalue $e_5$ is positive and yields consequently
an amplification of an infinitesimal perturbation.

For the ordered solution, there is no simple closed representation of the eigenvalues. However, we have checked that the numerical  
computation of the eigenvalues of $\cal J$ at the ordered solution $\omega=\omega_{\rm d}$ does imply linear stability
of the ordered solution in the region $g\in(g_1,g_2)$ and instability outside of this region.

\setcounter{footnote}{0} 

{\small 
\begin{thebibliography}{999}
\bibitem{Abra65} M. Abramowitz and I.A. Stegun, {\it Handbook of Mathematical Functions}, Dover (New York 1965)
\bibitem{Alta15} B. Altaner, A. Wachtel and J. Vollmer, Phys. Rev. {\bf E92}, 042133 (2015) {\tt [arxiv:1504.03648]}. 
\bibitem{Amit84} D.J. Amit and V. Mart\'{\i}n-Mayor, {\it Field theory, the renormalization group and critical phenomena},
3$^{\rm rd}$ ed., World Scientific (Singapour 1984, $^3$2005)
\bibitem{Atta06} S. Attal, A. Joyce and C.-A. Pillet, {\it Open quantum systems II: the markovian approach}, 
Springer Lecture Notes in Mathematics, LNM {\bf 1881}, Springer (Heidelberg 2006). 
\bibitem{Atta07} S. Attal and A. Joyce, J. Funct. Anal. {\bf 247}, 253 (2007) {\tt [arxiv:math-ph/0612055]}. 
\bibitem{Barm13} P. Barmettler, D. Fioretto and V. Gritsev, Europhys. Lett. {\bf 104}, 10004 (2013) {\tt [arxiv:1201.4416]}. 
\bibitem{Batc15} M.T. Batchelor and H.-Q. Zhou, Phys. Rev. {\bf A91}, 053808 (2015) {\tt [arxiv:1408.3816]}. 
\bibitem{Berg07} D. Bergman, J. Alicea, E. Gull, S. Trebst and L. Ballets, Nature Physics {\bf 3}, 487 (2007) {\tt [arXiv:cond-mat/0612001]}. 
\bibitem{Berl52} T.H. Berlin and M. Kac, Phys. Rev. {\bf 86}, 821 (1952).
\bibitem{Boas} M.L. Boas, {\it Mathematical methods in the physical sciences }, 2$^{\rm nd}$ ed., Wiley (New York 1983)
\bibitem{Borc14} N. Borchers, M. Pleimling and R.K.P. Zia, Phys. Rev. {\bf E90}, 062113 (2014) {\tt [arxiv:1411.6180]}
\bibitem{Braa13} D. Braak, J. Phys. {\bf B46}, 224007 (2013) {\tt [arxiv:1304.2529]}. 
\bibitem{Braa15} D. Braak, in Proc. of Forum of Mathematics for Industry 2014, {\tt [arxiv:1509.05748]}. 
\bibitem{Bran00} J.G. Brankov, D.M. Danchev and N.S. Tonchev, 
{\it Theory of critical phenomena in finite-size systems}, World Scientific (Singapour 2000). 
\bibitem{Breu02} H.-P. Breuer and F. Petruccione, {\it The theory of open quantum systems}, Oxford University Press (Oxford 2002). 
\bibitem{Card96} J.L. Cardy, {\it Scaling and renormalisation in statistical physics}, 
Cambridge University Press (Cambridge 1996). 
\bibitem{Carm99} H.J. Carmichael, {\it Statistical methods in quantum optics 1}, Springer (Heidelberg 1999).
\bibitem{Cham06} C. Chamon, L.F. Cugliandolo and H. Yoshino, J. Stat. Mech. P01006 (2006) {\tt [arxiv:cond-mat/0506297]}. 
\bibitem{Cugl95} L.F. Cugliandolo and D. Dean, J. Phys. {\bf A28}, 4213 (1995) {\tt [arxiv:cond-mat/9502075]}. 
\bibitem{Cugl03} L.F. Cugliandolo, in J.-L. Barrat, M. Feiglman, J. Kurchan, J. Dalibard (eds), 
{\it Slow relaxations and non-equilibrium dynamics in condensed matter}, Les Houches LXXVII, 
Springer (Heidelberg 2003), pp. 367-521 {\tt [arxiv:cond-mat/0210312]}. 
\bibitem{Dick54} R. H. Dicke, Phys. Rev. {\bf 93}, 99 (1954).
\bibitem{Dura15} X. Durang, C. Kwon and H. Park, Phys. Rev. {\bf E91}, 062118 (2015) {\tt [arxiv:1309.5750]}. 
\bibitem{Ehre10} D. Ehre, E. Lavert, M. Lahav and I. Lubomirsky, Science {\bf 327}, 672 (2010). 
\bibitem{Engl02} B.-G. Englert and G. Morigi, in A. Buchleitner and K. Hornberger (eds) 
{\it Coherent Evolution in Noisy Environments}, Springer Lecture Notes in Physics {\bf 611}, 
Springer (Heidelbereg 2002); pp. 55-106; {\tt [arxiv:quant-ph/0206116]}. 
\bibitem{Elou15} C. Elouard, A. Auff\`eves and M. Clusel, {\tt [arxiv:1507.00312]}. 
\bibitem{Fell71} W. Feller, {\it An introduction to probability theory and its applications}, vol. 2 (2$^{\rm nd}$ ed), 
Wiley (New York 1971). 
\bibitem{Fort12} J.-Y. Fortin and S. Mantelli, J. Phys. {\bf A45}, 475001 (2012) {\tt [arxiv:1208.2111]}. 
\bibitem{Fusc02} N. Fusco and M. Zannetti, Phys. Rev. {\bf E66}, 066113 (2002) {\tt [cond-mat/0210502]}. 
\bibitem{Fydo15} Y.V. Fyodorov, A. Perret and G. Schehr, J. Stat. Mech. P11017 (2015) {\tt [arxiv:1507.08520]}. 
\bibitem{Gara99} D.A. Garanin and B. Canals, Phys. Rev. {\bf B59}, 443 (1999) {\tt [arxiv:cond-mat/9805362]}. 
\bibitem{Garr11} B.M. Garraway, Phil. Trans. Roy. Soc. {\bf A369}, 1137 (2011). 
\bibitem{Godr00b} C. Godr\`eche and J.M. Luck, J. Phys. {\bf A33}, 9141 (2000) {\tt [arxiv:cond-mat/0001264]}.  
\bibitem{Godr02} C. Godr\`eche and J.-M. Luck, J. Phys. Cond. Matt. {\bf 14}, 1589 (2002) {\tt [arxiv:cond-mat/0109212]}. 
\bibitem{Godr13} C. Godr\`eche and J.-M. Luck, J. Stat. Mech. P05006 (2013) {\tt [arXiv:1302.4658]}. 
\bibitem{Gome13} P.R.S. Gomes, P.F. Bienzobaz and M. Gomes, Phys. Rev. {\bf D88}, 025050 (2013) {\tt [arxiv:1305.3792]}. 
\bibitem{Helb00} D. Helbing, I.J. Farkas and T. Vicsek, Phys. Rev. Lett. {\bf 84}, 1240 (2000) {\tt [arxiv:cond-mat/9904326]}. 
\bibitem{Henk84a} M. Henkel and C. Hoeger, Z. Phys. {\bf B55}, 67 (1984). 
\bibitem{Henk09} M. Henkel, H. Hinrichsen and S. L\"ubeck, 
{\it ``Non-equilibrium phase transitions vol. 1: absorbing phase transitions''}, Springer (Heidelberg 2009).
\bibitem{Henk10} M. Henkel and M. Pleimling, 
{\it ``Non-equilibrium phase transitions vol. 2: 
ageing and dynamical scaling far from equilibrium''}, Springer (Heidelberg 2010).
\bibitem{Henk15} M. Henkel and X. Durang, J. Stat. Mech. P05022 (2015) {\tt [arxiv:1501.07745]}. 
\bibitem{Hol40} T. Holstein and H. Primakoff, Phys. Rev. {\bf 58 }, 1098 (1940)
\bibitem{Isak04} S.V. Isakov, K. Gregor, R. Moessner and S.L. Sondhi, 
Phys. Rev. Lett. {\bf 93}, 167204 (2004) {\tt [arxiv:cond-mat/0407004]}. 
\bibitem{Isak15} S.V. Isakov, R. Moessner, S.L. Sondhi and D.A. Tennant, Phys. Rev. {\bf B91}, 245152 (2015) {\tt [arxiv:1504.04156]}. 
\bibitem{Jaeg96} E. J\"ager and R. Perthel, {\em Magnetische Eigenschaften von Festk\"orpern}, Akademie Verlag (Berlin 1996). 
\bibitem{Joyc72} G.S. Joyce, in C. Domb, M.S. Green (eds), 
{\it Phase transitions and critical phenomena, vol. 2}, Academic Press (London 1972), p. 375.  
\bibitem{Kare13} D. Karevski, V. Popkov and G.M. Sch\"utz, Phys. Rev. Lett. {\bf 110}, 047201 (2013) {\tt [arxiv:1211.7010]}. 
\bibitem{Katz84} S. Katz, J.L. Lebowitz and H. Spohn, J. Stat. Phys. {\bf 34}, 497 (1984). 
\bibitem{Lewi52} H.W. Lewis and G.H. Wannier, Phys. Rev. {\bf 88}, 682 (1952); erratum {\bf 90}, 1131 (1953). 
\bibitem{Niew95} Th. M. Nieuwenhuizen, Phys. Rev. Lett. {\bf 74}, 4293 (1995) {\tt [arxiv:cond-mat/9408056]}. 
\bibitem{Nish11} H. Nishimori and G. Ortiz, {\it Elements of phase transitions and critical phenomena}, 
Oxford University Press (Oxford 2011). 
\bibitem{Ober72} G. Obermair, in J.I. Budnick and M.P. Kawars (eds), 
{\it  Dynamical Aspects of Critical Phenomena}, (Gordon and Breach, New York, 1972), p. 137. 
\bibitem{Oliv06} M. H. Oliveira, E.P. Raposo and M.D. Coutinho-Filho, Phys. Rev. {\bf B74}, 184101 (2006). 
\bibitem{Podo14} D. Podolsky, E. Shimshoni, P. Silvi, S.  Montangero, T. Calarco, G. Morigi, S.  Fishman, 
Phys. Rev. {\bf B89}, 214408 (2014) {\tt [arxiv:1403.2422]}. 
\bibitem{Poly} A.D. Polyanin and V.F. Zaitsev, {\it Handbook of Exact Solutions for Ordinary Differential Equations}, 
2$^{\rm nd}$ ed., Chapman \& Hall (London 2002). 
\bibitem{Popk13} V. Popkov, D. Karevski, and G.M. Sch\"utz, Phys. Rev. {\bf E88}, 062118 (2013) {\tt [arxiv:1310.1315]}. 
\bibitem{Pros11} T. Prosen, Phys. Rev. Lett. {\bf 106}, 217206 (2011) {\tt [arxiv:1103.1350]}. 
\bibitem{Pros15} T. Prosen, J. Phys. {\bf A48}, 373001 (2015) {\tt [arxiv:1504.00783]}. 
\bibitem{Radz03} L. Radzihovsky and N.A. Clark, Phys. Rev. Lett. {\bf 90}, 189603 (2003)
\bibitem{Ronc78} G. Ronca, J. Chem. Phys. {\bf 68}, 3737 (1978). 
\bibitem{Ross14} K.A. Ross, Y. Qiu, J.R.D. Copley, H.A. Dabkowska and B.D. Gaulin, 
Phys. Rev. Lett. {\bf 112}, 057201 (2014) {\tt [arxiv:1401.1176]}. 
\bibitem{Ross12} D.Z. Rossatto, A.R. de Almeida, T. Werlang, C.J. Villas-Boas and N.G. de Almeida, 
Phys. Rev. {\bf A86}, 035802 (2012) {\tt [arXiv:1210.0239]}. 
\bibitem{Sach99} S. Sachdev, {\it Quantum phase transitions}, Cambridge University Press (Cambridge 1999). 
\bibitem{Scha14} G. Schaller, {\it Open quantum systems far from equlibrium}, Springer Lecture Notes in Physics LNP {\bf 881}, 
Springer (Heidelberg 2014). 
\bibitem{Shuk81} P. Shukla and S. Singh, Phys. Lett. {\bf 81A}, 477 (1981); Phys. Rev. {\bf B23}, 4661 (1981). 
\bibitem{Taeu14} U.C. T\"auber, {\it Critical dynamics}, Cambridge University Press (Cambridge 2014). 
\bibitem{Tail06} J. Tailleur, S. Tanase-Nicola and J. Kurchan, J. Stat. Phys. {\bf 122}, 557 (2006) {\tt [arxiv:cond-mat/0503545]}. 
\bibitem{Ton09}  N.S. Tonchev, J.G. Brankov, V.A. Zagrebnov, J. Optoel.  Adv. Mat. {\bf 11}, 1142 (2008). 
\bibitem{vanH15} M. van Horssen and J.P. Garrahan, Phys. Rev. {\bf E91}, 032132 (2015) {\tt [arxiv:1411.7913]}. 
\bibitem{Vill80} J. Villain, R. Bideaux, J.P. Carton and R. Conte, J. Physique {\bf 41}, 1263 (1980). 
\bibitem{Vojt96} T. Vojta, Phys. Rev. {\bf B53}, 710 (1996). 
\bibitem{Wald15} S. Wald and M. Henkel, J. Stat. Mech. P07006 (2015) {\tt [arxiv:1503.06713]}. 
\bibitem{Wald16} S. Wald, G.T. Landi and M. Henkel, {\it in preparation}
\bibitem{Wipf13} A. Wipf, {\it Statistical approach to quantum-field-theory}, Springer Lecture Notes in Physics {\bf 864}, 
Springer (Heidelberg 2013).
\bibitem{Zia02} R.K.P. Zia, E.L. Praestgaard and O.G. Mouritsen, Am. J. Phys. {\bf 70}, 384 (2002) {\tt [arxiv:cond-mat/018502]}. 
%

\end{thebibliography} }
\end{document}